\documentclass[12pt]{article}
\usepackage[a4paper, margin = 1in]{geometry}
\usepackage[sorting = none, maxbibnames=9, backend = bibtex, style=numeric-comp]{biblatex}
\usepackage[titletoc,page]{appendix}
\renewcommand*\appendixpagename{Appendix}
\renewcommand*\appendixtocname{Appendix}
\renewcommand*{\bibfont}{\footnotesize}
\renewbibmacro{in:}{}
\usepackage{graphicx}
\usepackage{setspace}
\usepackage{gensymb}
\usepackage{enumerate}
\usepackage{kotex}
\usepackage{abstract,lipsum}
\usepackage[colorlinks=true,linkcolor=blue,citecolor=blue,urlcolor=blue]{hyperref}
\usepackage{amsmath}
\usepackage{amsthm}
\usepackage{amsfonts}
\usepackage{upgreek}
\usepackage{mathtools}
\usepackage{mathrsfs}
\usepackage{bm}
\newcommand\defeq{\stackrel{\mathclap{\normalfont\tiny\mbox{def}}}{=}}
\usepackage[symbol]{footmisc}

\usepackage{color}

\begin{document}
\begin{center}
       \fontsize{19pt}{19pt}\selectfont A polyurethane-urea elastomer at low to extreme strain rates
       
       \vspace*{0.3in}
       \fontsize{10pt}{10pt}\selectfont Jaehee Lee$^{1}$, David Veysset$^{2,3}$, Alex J. Hsieh$^{2,4}$, Gregory C. Rutledge$^{5}$, Hansohl Cho$^{1,\dagger}$\\
       \vspace*{0.3in}
       \fontsize{9pt}{9pt}\selectfont {$^{1}$ Department of Aerospace Engineering, Korea Advanced Institute of Science and Technology, Daejeon, 34141, Republic of Korea \\ 
       $^{2}$ Institute for Soldier Nanotechnologies, Massachusetts Institute of Technology, Cambridge, MA 02139, USA \\ 
       $^{3}$ Hansen Experimental Physics Laboratory, Stanford University, Stanford, CA 94305, USA \\ 
       $^{4}$ U.S. Army Combat Capabilities Development Command, Army Research Laboratory, FCDD-RLW-MG, Aberdeen Proving Ground, MD 21005-5069, USA \\   
       $^{5}$ Department of Chemical Engineering, Massachusetts Institute of Technology, Cambridge, MA 02139, USA} \\
\vspace*{0.1in}
$^\dagger$E-mail: hansohl@kaist.ac.kr
\end{center}

\renewenvironment{abstract}
{\small 
\noindent \rule{\linewidth}{.5pt}\par{\noindent \bfseries \abstractname.}}
{\medskip\noindent \rule{\linewidth}{.5pt}
}

\vspace*{0.3in}
\onehalfspacing
\begin{abstract}
\fontsize{10pt}{10pt}\selectfont
A finite strain nonlinear constitutive model is presented to study the extreme mechanical behavior of a polyurethane-urea well suited for many engineering applications. The micromechanically- and thermodynamically based constitutive model captures salient features in resilience and dissipation in the material from low to extreme strain rates. The extreme deformation features are further elucidated by laser-induced micro-particle impact tests for the material, where an ultrafast strain rate ($>10^6$ s$^{-1}$) incurs. Numerical simulations for the strongly inhomogeneous deformation events are in good agreement with the experimental data, supporting the predictive capabilities of the constitutive model for the extreme deformation features of the PUU material over at least 9 orders of magnitude in strain rates ($10^{-3}$ to $10^{6}$ s$^{-1}$). \\
\end{abstract} 

\doublespacing
\section{Introduction}
Polyurethane-urea (PUU) materials are phase-separated copolymers that possess remarkable combinations of mechanical properties of elastomeric and plastomeric materials. The hybrid mechanical features involving both shape recovery and energy dissipation in the PUU materials result from the phase-separating microstructures of hard and soft domains which are usually below and above glass transition, respectively, at ambient conditions. Such hybrid mechanical features in the PUU materials have facilitated their applications in multi-functional soft material architectures ranging from stimuli-responsive polymeric materials to soft robotics and actuation, bio-compatible devices, and impact- or shock-mitigating composites, often subjected to extreme strains and strain rates.

Over the past two decades, this important class of two-phase elastomers has stimulated active experimental and theoretical research on their versatile mechanical responses at a wide range of strains and strain rates. Specifically, the highly nonlinear rate-dependent elastic-inelastic stress-strain behaviors in both tension and compression have been widely reported, where multiple dissipation mechanisms involving viscoelasticity and viscoplasticity were elucidated \cite{Yi2006,Rinaldi2010,Rinaldi2011}. Furthermore, the stretch-induced softening also known as the Mullins’ effect has been extensively investigated in these materials via multiple cyclic loading and unloading tests and \textit{in situ} small- and wide-angle X-ray scattering measurements (SAXS/WAXS) \cite{Rinaldi2010,Stribeck2017}, by which the microstructural breakdown in hard domains was found to provide another major energy dissipation pathway. Also, the high strain rate responses ($> 10^{3}$ s$^{-1}$) have been widely studied in the split Hopkinson pressure bar (SHPB) setup and the Taylor-type gas-gun apparatus \cite{Sarva2007,Roland2007}. In conjunction with experimental studies, there have been theoretical and computational modeling efforts that aim at exploring the large strain mechanical behavior and the phase behavior of the hard and soft components in the polyurethane and polyurea materials at molecular- \cite{Heyden2016,Rutledge2017,Brinson2018} and continuum scales \cite{Cho2013a,Nemat2006,Sayed2009,Clifton2016}.

While there is vast literature on the mechanical behavior of individual polyurethane or polyurea materials across a wide range of length-scales, thermodynamically- and microstructurally-based constitutive models for the PUU materials having both urethane- and urea contents valid from very low to extremely high strain rates are largely lacking. In this work, we address the extreme mechanical behaviors of a representative PUU material (41wt\% for hard; 59wt\% for soft) by constitutive modeling, experiments, and numerical simulations. We present a thermodynamically- and microstructurally-based nonlinear constitutive model able to capture the salient features of the large strain behaviors of the exemplar PUU material in a broad range of loading conditions including both tension and compression up to extreme strains.


Then, the extreme deformation behavior of the PUU material is further elucidated in micro-sized particle impact experiments, where an ultrafast strain rate ($> 10^{6}$ s$^{-1}$) incurs in conjunction with a strongly inhomogeneous deformation field. To this end, we conducted a laser-induced particle impact test (LIPIT) to elucidate both resilience and dissipation in the PUU material under the extreme deformation event on a timescale of hundreds of nanoseconds. Last, we performed numerical simulations for the LIPIT of the PUU material for which the proposed nonlinear constitutive model was numerically implemented. We then compared the numerical results against the corresponding experimental data collected in the laser-induced micro-particle impact test, by which we show that our modeling strategy is capable of capturing the extreme deformation features of this important class of elastomeric (or plastomeric) materials over at least 9 orders of magnitude in strain rates (10$^{-3}$ to 10$^{6}$ s$^{-1}$) and up to an extreme strain of $> 600\%$.

\section{Constitutive model}
\begin{figure}[b!]
\includegraphics[width=1.0\textwidth]{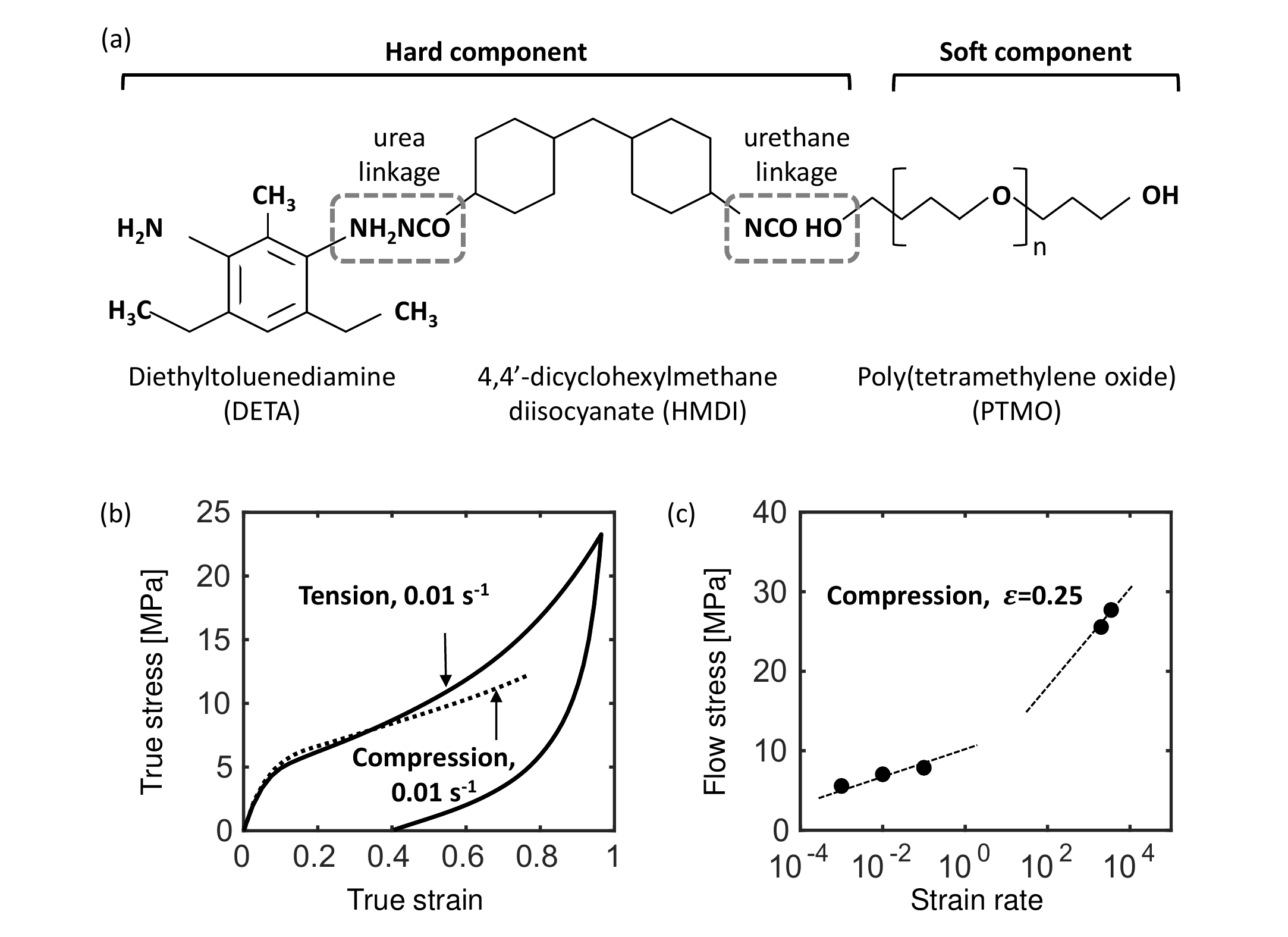} 
\caption{(a) Schematic illustration of the chemical structure of the PUU material with urea- and urethane-linkages. (b) Stress vs. strain curves  at a strain rate of 0.01 s$^{-1}$ in tension and compression (up to a strain of 0.8) and (c) flow stress vs. strain rate at a strain of 0.25 in compression. \\}
\label{fig:PUUchemical}
\end{figure}
The PUU material discussed in this work is composed of 4,4'-dicyclohexyl-methane diisocyanate (HMDI) with chain extender diethyltoluenediamine (DETA) for hard domains and poly(tetramethylene oxide) (PTMO) with a molecular weight of 2000 g/mol for soft domains (41wt\% hard contents). As illustrated in Figure \ref{fig:PUUchemical} (a), the hard and soft components of the PUU material are connected via the urethane- and the urea-linkages, often forming a phase-separated morphology \cite{Rinaldi2010,Runt2000,Runt2001}. Detailed information on the physical and chemical properties and the synthesis procedures for the material can be found in Rinaldi et al \cite{Rinaldi2010}, where the large strain mechanical tests were conducted in conjunction with \textit{in situ} SAXS and WAXS measurements for monitoring the microstructural evolution during deformation. Here, we present a three-dimensional constitutive modeling framework for this representative PUU material.

Numerous constitutive models exist in the literature for polyurethane and polyurea materials with a wide range of weight fractions of hard and soft contents. Some early models captured the low strain rate compressive behavior of the materials without any considerations on stretch-induced softening \cite{Nemat2006,Shim2011}. Stretch-induced softening was then captured for polyurethane materials by Qi and Boyce \cite{Qi2004}. Their ideas on phase conversion upon deformation have found success in modeling the Mullins’ effect in other two-phase materials \cite{Wang2018a,Wang2018b}. Microstructurally-based models \cite{Cho2017} have also found success in capturing the softening, for which a simple damage model based on the theory of network breakdown and alteration \cite{Marckmann2002,Chagnon2006} was introduced. Turning attention to the high strain rate behavior, some modeling efforts \cite{Sayed2009,Cho2013b,Clifton2016} have shown capabilities of simulating the extreme deformation events in the materials incurred during the Taylor gas gun tests and the pressure-shear impact tests. However, most of the existing models only attempted to predict either quasi-static low strain rate behavior or high strain rate behavior. Moreover, these early models have not considered an appropriate thermodynamic framework for the constitutive modeling procedures.

Herein, our overall goal is to present a thermodynamically- and microstructurally-based nonlinear constitutive modeling framework for the exemplar polyurethane-urea elastomer valid at an extremely wide range of strains and strain rates under homogeneous and inhomogeneous deformation events. Specifically, the proposed constitutive model should be able to capture the following major mechanical features in both tension and compression at very large strains: (1) highly nonlinear elastic-inelastic stress-strain behavior accompanied by remarkable hysteresis and shape recovery upon loading and unloading, (2) strong asymmetry between tension and compression especially at large strains, as shown in Figure \ref{fig:PUUchemical} (b), (3) rate-dependent yields and flow stresses and the change in the rate-sensitivity\footnote{The change in the rate-sensitivity due to the distinct relaxation processes in hard and soft domains is further supported by the dynamic mechanical analysis results on this representative PUU material reported in Rinaldi et al \cite{Rinaldi2010}.} due to the presence of the two distinct domains of hard and soft components, as shown in Figure \ref{fig:PUUchemical} (c), and (4) stretch-induced softening (Mullins’ effect) due to the microstructural breakdown especially in hard domains within an appropriate thermodynamic framework.


Our constitutive model comprises four micro-rheological mechanisms assumed to act in parallel to account for the stress resistances from each of the hard and soft domains. For the hard domains, we employed a time-dependent hyperelastic-viscoplastic mechanism and a time-independent hyperelastic mechanism denoted by ``h$_{1}$" and ``h$_{2}$", respectively. Here, the h$_{1}$ mechanism represents a resistance to the internal energy change due to intermolecular interactions such as chain-segment rotation and slippage between the polymer chains. The h$_{2}$ mechanism represents a resistance to the entropic free energy change upon stretching and orientation of the polymer chains. The micro-mechanisms in soft domains are almost identical to those in hard domains. Here, ``s$_{1}$" and ``s$_{2}$" mechanisms represent the resistances to the internal energy change and entropy change in soft domains, respectively.

\subsection{Kinematics}
\noindent A motion $\upvarphi$ is defined as a one-to-one mapping $\mathbf{x}=\upvarphi(\mathbf{X},t)$ with a material point $\mathbf{X}$ in a fixed undeformed reference and $\mathbf{x}$ in a deformed spatial configuration. The large deformation kinematics for the elastic-inelasticity follows a multiplicative decomposition of a deformation gradient $\mathbf{F}\defeq\frac{\partial\upvarphi}{\partial\mathbf{X}}$,
\begin{equation}
\mathbf{F}=\mathbf{F}^{e(\mathrm{h_{1}})}\mathbf{F}^{p(\mathrm{h_{1}})}=\mathbf{F}^{(\mathrm{h_{2}})}=\mathbf{F}^{e(\mathrm{s_{1}})}\mathbf{F}^{p(\mathrm{s_{1}})}=\mathbf{F}^{e(\mathrm{s_{2}})}\mathbf{F}^{p(\mathrm{s_{2}})}
\end{equation}

\noindent
where $\mathbf{F}^{e(\alpha)}$ and $\mathbf{F}^{p(\alpha)}$ are the elastic and inelastic deformation gradients for the time-dependent mechanisms $\alpha=$ h$_{1}$, s$_{1}$, and s$_{2}$.  $\mathbf{F}$ allows for polar decomposition, $\mathbf{F}=\mathbf{R}\mathbf{U}=\mathbf{V}\mathbf{R}$ with the rotation $\mathbf{R}$, and the right and left stretch tensors $\mathbf{U}$ and $\mathbf{V}$. Then, we define,
\begin{equation}
\begin{aligned}
& \text{polar decomposition of } \, \mathbf{F}^{e} \qquad && \mathbf{F}^{{e(\alpha)}}=\mathbf{R}^{{e(\alpha)}}\mathbf{U}^{e(\alpha)}= \mathbf{V}^{{e(\alpha)}}\mathbf{R}^{e(\alpha)} \\
& \text{elastic right Cauchy-Green tensor} \qquad && \mathbf{C}^{e(\alpha)}=\mathbf{F}^{{e(\alpha)}\top}\mathbf{F}^{e(\alpha)} \\
& \text{elastic left Cauchy-Green tensor} && \mathbf{B}^{e(\alpha)}=\mathbf{F}^{e(\alpha)}\mathbf{F}^{{e(\alpha)}\top} \\
& \text{isochoric part of } \, \mathbf{F}^{e} && \bar{\textbf{F}}^{e(\alpha)}=(J^{e(\alpha)})^{-1/3}\textbf{F}^{e(\alpha)}, \quad J^{e(\alpha)}\defeq\det\textbf{F}^{e(\alpha)} \\
& \text{isochoric part of } \, \mathbf{C}^{e} && \Bar{\textbf{C}}^{e(\alpha)}=\bar{\mathbf{F}}^{{e(\alpha)}\top}\bar{\mathbf{F}}^{e(\alpha)} \\
& \text{isochoric part of } \, \mathbf{B}^{e} && \Bar{\textbf{B}}^{e(\alpha)}=\bar{\mathbf{F}}^{e(\alpha)}\bar{\mathbf{F}}^{{e(\alpha)}\top} \\
\end{aligned}
\end{equation}

\noindent The deformation rate is described by the velocity gradient $\mathbf{L}\defeq\mathrm{grad}\textbf{v}=\frac{\partial \textbf{v}}{\partial \textbf{x}}$ decomposed into elastic and inelastic parts,
\begin{equation}
\begin{aligned}
\mathbf{L}^{(\alpha)} & =\dot{\mathbf{F}}^{(\alpha)}\mathbf{F}^{(\alpha)-1} \\
& =\dot{\mathbf{F}}^{e(\alpha)}\mathbf{F}^{e(\alpha)-1}+\mathbf{F}^{e(\alpha)}\dot{\mathbf{F}}^{p(\alpha)}\mathbf{F}^{p(\alpha)-1}\mathbf{F}^{e(\alpha)-1} \\
& = \mathbf{L}^{e(\alpha)}+\mathbf{F}^{e(\alpha)}\mathbf{L}^{p(\alpha)}\mathbf{F}^{e(\alpha)-1}.
\end{aligned}
\end{equation}

\noindent The symmetric part of $\mathbf{L}^{p(\alpha)}$ is the inelastic stretching, $\mathbf{D}^{p(\alpha)}$; and the skew part of $\mathbf{L}^{p(\alpha)}$ is the inelastic spin, $\mathbf{W}^{p(\alpha)}$; i.e., $\mathbf{L}^{p(\alpha)}=\mathbf{D}^{p(\alpha)}+\mathbf{W}^{p(\alpha)}$. Furthermore, we make two important kinematical assumptions for the inelastic flow; the flow is incompressible, i.e. $J^{p(\alpha)}\defeq\det\mathbf{F}^{p(\alpha)}=1$ and irrotational, i.e. $\mathbf{W}^{p(\alpha)}=0$. Thus, the rate of inelastic deformation gradient is given by
\begin{equation}
\begin{aligned}
\Dot{\mathbf{F}}^{p(\alpha)} = \mathbf{D}^{p(\alpha)}\mathbf{F}^{p(\alpha)}. 
\label{eq:Fpderivative}
\end{aligned}
\end{equation}

\subsection{Thermodynamic framework}
We consider the first and second laws of thermodynamics for the constitutive model. Combining the first two laws of thermodynamics, we have a free energy imbalance for each part P of the reference body that takes the form of, 
\begin{equation}
\int_{\mathrm{P}} \left( \dot{\psi}_{\mathrm{R}} - \left(\sum_{\mathrm{h_{1},s_{1},s_{2}}}\left(\mathbf{S}^{e(\alpha)}:\dot{\mathbf{F}}^{e(\alpha)} + \mathbf{T}^{p(\alpha)}:\mathbf{D}^{p(\alpha)}\right) + \mathbf{S}^{(\mathrm{h_{2}})}:\dot{\mathbf{F}} \right)\right)dv_{\mathrm{R}} \leq 0 
\label{eq:FrEnerImb}
\end{equation}
\noindent with the elastic stresses $\mathbf{S}^{e(\alpha)}$ conjugate to $\dot{\mathbf{F}}^{e(\alpha)}$ and the inelastic stresses $\mathbf{T}^{p(\alpha)}$ conjugate to $\mathbf{D}^{p(\alpha)}$ for $\alpha=$ h$_{1}$, s$_{1}$, and s$_{2}$ mechanisms, and the elastic stress $\mathbf{S}^{(\mathrm{h_{2}})}$ conjugate to $\dot{\mathbf{F}}$. We then define,
\begin{equation}
\begin{aligned}
& \text{first Piola stress } \qquad && \mathbf{T}_{\mathrm{R}}\defeq \sum_{\mathrm{h_{1},s_{1},s_{2}}}\mathbf{S}^{e(\alpha)}\mathbf{F}^{p(\alpha)-\top} + \mathbf{S}^{(\mathrm{h_{2}})} \\
& \text{second Piola stress } \qquad && \mathbf{T}^{(\mathrm{h_{2}})}_{\mathrm{RR}} \defeq J\mathbf{F}^{-1}\mathbf{T}^{(\mathrm{h_{2}})}\mathbf{F}^{-\top} \\
& \text{elastic second Piola stress for } \alpha=\text{h}_{1},\text{s}_{1},\text{s}_{2} \qquad && \mathbf{T}^{e(\alpha)} \defeq J^{e(\alpha)}\mathbf{F}^{e(\alpha)-1}\mathbf{T}^{(\alpha)}\mathbf{F}^{e(\alpha)-\top} \\
& \text{Mandel stress for } \alpha=\text{h}_{1},\text{s}_{1},\text{s}_{2} \qquad && \mathbf{M}^{e(\alpha)} \defeq \mathbf{C}^{e(\alpha)}\mathbf{T}^{e(\alpha)} \\
\end{aligned}
\end{equation}
where the first Piola stress $\mathbf{T}_{\mathrm{R}}$ is related to the Cauchy stress $\mathbf{T}$ by $\mathbf{T}_{\mathrm{R}}=J\mathbf{T}\mathbf{F}^{-\top}$. Therefore, we have,
\begin{equation}
\begin{aligned}
& \mathbf{T} = \sum_{\mathrm{h_{1},s_{1},s_{2}}} J^{-1}\mathbf{S}^{e(\alpha)}\mathbf{F}^{e(\alpha)\top} + J^{-1}\mathbf{S}^{(\mathrm{h_{2}})}\mathbf{F}^{\top} \\
\text{with} \qquad & \mathbf{T}^{(\alpha)} = J^{-1}\mathbf{S}^{e(\alpha)}\mathbf{F}^{e(\alpha)\top} \qquad \text{and} \qquad \mathbf{T}^{(\mathrm{h_{2}})} = J^{-1}\mathbf{S}^{(\mathrm{h_{2}})}\mathbf{F}^{\top} \\
\end{aligned}
\end{equation}
Since the inelastic flow is assumed to be incompressible, i.e., $J^{e(\alpha)}=J$, we obtain $\mathbf{T}^{e(\alpha)}=\mathbf{F}^{e(\alpha)-1}\mathbf{S}^{e(\alpha)}$. Hence, the elastic stress power for $\alpha=$ h$_{1}$, s$_{1}$, and s$_{2}$ mechanisms can be expressed by,
\begin{equation}
\begin{aligned}
\mathbf{S}^{e(\alpha)}:\dot{\mathbf{F}}^{e(\alpha)} & =\mathbf{T}^{e(\alpha)}:\left(\mathbf{F}^{e(\alpha)\top}\dot{\mathbf{F}}^{e(\alpha)}\right) \\
& = \frac{1}{2}\mathbf{T}^{e(\alpha)}:\dot{\mathbf{C}}^{e(\alpha)}
\end{aligned}
\end{equation}
\noindent Similarly, using $\mathbf{T}^{(\mathrm{h_{2}})}_{\mathrm{RR}}=\mathbf{F}^{-1}\mathbf{S}^{(\mathrm{h_{2}})}$,
\begin{equation}
\begin{aligned}
\mathbf{S}^{(\mathrm{h_{2}})}:\dot{\mathbf{F}} & =\mathbf{T}^{(\mathrm{h_{2}})}_{\mathrm{RR}}:\mathbf{F}^{\top}\dot{\mathbf{F}} = \frac{1}{2}\mathbf{T}^{(\mathrm{h_{2}})}_{\mathrm{RR}}:\dot{\mathbf{C}}. \\
\end{aligned}
\end{equation}
Furthermore, from the microscopic force balance that characterizes the interaction between internal forces associated with the elastic and inelastic responses in each of the micromechanisms \cite{Anand2009, Mao2017}, we have, $\mathbf{M}^{e(\alpha)}_{0}=\mathbf{T}^{p(\alpha)}$ where the deviatoric part of the Mandel stresses $\mathbf{M}^{e(\alpha)}_{0}=\mathbf{M}^{e(\alpha)} - 1/3(\text{tr}\mathbf{M}^{e(\alpha)})$.\footnote{From the microforce balance, we have $\mathrm{sym}\mathbf{M}^{e(\alpha)}_{0}=\mathbf{T}^{p(\alpha)}$. Then, since the free energy functions are assumed to be isotropic in this work, $\mathbf{C}^{e(\alpha)}$ and $\mathbf{T}^{e(\alpha)}$ commute; i.e. $\mathbf{C}^{e(\alpha)}\mathbf{T}^{e(\alpha)}=\mathbf{T}^{e(\alpha)}\mathbf{C}^{e(\alpha)}$. Thus, $\mathbf{M}^{e(\alpha)}$ is symmetric due to the elastic isotropy.} Thus, the free energy imbalance in Equation (\ref{eq:FrEnerImb}) can be rewritten as,
\begin{equation}
\int_{\mathrm{P}} \left( \dot{\psi}_{\mathrm{R}} - \sum_{\mathrm{h_{1},s_{1},s_{2}}}\frac{1}{2}\mathbf{T}^{e(\alpha)}:\dot{\mathbf{C}}^{e(\alpha)} - \frac{1}{2}\mathbf{T}^{(\mathrm{h_{2}})}_{\mathrm{RR}}:\dot{\mathbf{C}}^{(\mathrm{h_{2}})} - \sum_{\mathrm{h_{1},s_{1},s_{2}}}\mathbf{M}^{e(\alpha)}_{0}:\mathbf{D}^{p(\alpha)} \right)dv_{\mathrm{R}} \leq 0. \\ 
\label{eq:FrEnerImblocal}
\end{equation}
Here, we assume the free energy, $\psi_{\mathrm{R}}$ in reference to be the form of,
\begin{equation}
\psi_{\mathrm{R}}=\hat\psi_{\mathrm{R}}^{(\mathrm{h_{1}})}(\mathbf{C}^{e(\mathrm{h_{1}})})+\hat\psi_{\mathrm{R}}^{(\mathrm{h_{2}})}(\mathbf{C},\xi)+\hat\psi_{\mathrm{R}}^{(\mathrm{s_{1}})}(\mathbf{C}^{e(\mathrm{s_{1}})})+\hat\psi_{\mathrm{R}}^{(\mathrm{s_{2}})}({\mathbf{C}}^{e(\mathrm{s_{2}})})
\label{eq:FreeE}
\end{equation}

\noindent
with an internal variable $\xi$ that represents the microstructural damage in the h$_{2}$ mechanism detailed in Section 2.3. 
Substituting the free energy form in Equation (\ref{eq:FreeE}) into the free energy imbalance in Equation (\ref{eq:FrEnerImblocal}), the local form of the free energy imbalance can be written as,
\begin{equation}
\begin{aligned}
\sum_{\mathrm{h_{1},s_{1},s_{2}}} & \left(\frac{\partial\hat\psi_{\mathrm{R}}^{(\alpha)}\left(\mathbf{C}^{e(\alpha)}\right)}{\partial\mathbf{C}^{e(\alpha)}}-\frac{1}{2}\mathbf{T}^{e(\alpha)}(\mathbf{C}^{e(\alpha)})\right):\dot{\mathbf{C}}^{e(\alpha)} \quad - \quad \sum_{\mathrm{h_{1},s_{1},s_{2}}} \mathbf{M}^{e(\alpha)}_{0}:\mathbf{D}^{p(\alpha)} \\ \\
+ \quad & \left(\frac{\partial\hat\psi_{\mathrm{R}}^{(\mathrm{h_{2}})}\left(\mathbf{C}, \xi\right)}{\partial\mathbf{C}}-\frac{1}{2}\mathbf{T}_{\mathrm{RR}}^{(\mathrm{h_{2}})}(\mathbf{C}, \xi)\right):\dot{\mathbf{C}} \quad + \quad \frac{\partial\hat\psi_{\mathrm{R}}^{(\mathrm{h_{2}})}\left(\mathbf{C}, \xi\right)}{\partial\xi}\dot{\xi} \,\leq 0. \\
\end{aligned}
\label{eq:thermorest}
\end{equation}
From the thermodynamic restrictions, we have, 
\begin{equation}
\begin{aligned}
& \mathbf{T}^{e(\alpha)} =  2\frac{\partial\hat{\psi}_{\mathrm{R}}^{(\alpha)}(\mathbf{C}^{e(\alpha)})}{\partial \mathbf{C}^{e(\alpha)}} \quad \text{for} \quad \alpha=\mathrm{h_{1},s_{1},s_{2}} \\
& \mathbf{T}_{\mathrm{RR}}^{(\mathrm{h_{2}})} =  2\frac{\partial\hat{\psi}_{\mathrm{R}}^{(\mathrm{h_{2}})}(\mathbf{C}, \xi)}{\partial \mathbf{C}}. \\
\end{aligned}
\label{eq:stressthermorest}
\end{equation}

\noindent Therefore, the following dissipation inequality must hold
\begin{equation}
 \mathcal{D} = \sum_{\mathrm{h_{1},s_{1},s_{2}}} \mathbf{M}^{e(\alpha)}_{0}:\mathbf{D}^{p(\alpha)} -\frac{\partial\hat\psi_{\mathrm{R}}^{(\mathrm{h_{2}})}\left(\mathbf{C}, \xi\right)}{\partial\xi}\dot{\xi} \, \geq 0.
\end{equation}
\noindent Here, the first terms are the contributions from the inelastic flows in the h$_{1}$, s$_{1}$ and s$_{2}$ mechanisms and the second term is the contribution from the microstructural damage in the h$_{2}$ mechanism. The dissipation inequalities must hold individually, i.e.,
\begin{equation}
\sum_{\mathrm{h_{1},s_{1},s_{2}}} \mathbf{M}^{e(\alpha)}_{0}:\mathbf{D}^{p(\alpha)}  \geq 0, \qquad -\frac{\partial\hat\psi_{\mathrm{R}}^{(\mathrm{h_{2}})}\left(\mathbf{C}, \xi\right)}{\partial\xi}\dot{\xi} \geq 0.
\label{eq:dissipation}
\end{equation}
The individual dissipation inequalities in Equation (\ref{eq:dissipation}) for each of the micromechanisms will be detailed in Section 2.3.

\subsection{Constitutive equations}

\subsubsection{Hard domain}
The constitutive equations for the time-dependent elastic-viscoplastic mechanism, h$_{1}$, and the time-independent hyperelastic mechanism, h$_{2}$, in the hard domain are presented. The h$_{1}$ mechanism captures a relatively stiff initial elastic response followed by a stress rollover for which a thermally-activated, rate-dependent viscoplastic flow model is employed. The h$_{2}$ mechanism captures the dramatic stress hardening at large strains as the deformation approaches the limiting chain extensibility in the molecular network. Furthermore, in the h$_{2}$ mechanism, we employ a simple damage process to account for the stretch-induced softening (Mullins' effect) widely observed in the two-phase polyurethane, polyurea and polyurethane-urea materials in cyclic loading conditions.
\paragraph{Intermolecular component ($\alpha=\mathrm{h_{1}}$)} \mbox{}\\
\noindent 
The free energy function $\psi_{\mathrm{R}}^{(\mathrm{h_{1}})}$ for the h$_{1}$ mechanism is taken to be a form of the three-dimensional linear elasticity given by 
\begin{equation}
\psi_{\mathrm{R}}^{(\mathrm{\mathrm{h_{1}}})} =\mu^\mathrm{h_{1}}|{\mathbf{E}^{e(\mathrm{h_{1}})}_{0}}|^{2}+\frac{1}{2}K(\text{tr}\mathbf{E}^{e(\mathrm{h_{1}})})^2
\label{eq:h1energy}
\end{equation}

\noindent with the shear modulus $\mu^\mathrm{h_{1}}$, the bulk modulus $K$, a logarithmic elastic strain (Hencky's strain \cite{Anand1979}) $\mathbf{E}^{e}=\ln{\mathbf{U}^{e}}$ and its deviatoric part $\mathbf{E}^{e}_{0}=\mathbf{E}^{e} - 1/3(\text{tr}\mathbf{E}^{e})$. 
It should be noted that the bulk response is lumped into the h$_{1}$ mechanism. 
The Cauchy stress is related to the Mandel stress by, 
\begin{equation}
\begin{aligned}
\mathbf{M}^{e(\mathrm{h_{1}})} &  = 2\mu^\mathrm{h_{1}}\mathbf{E}^{e(\mathrm{h_{1}})}_{0}+K(\mathrm{tr}\mathbf{E^{e(\mathrm{h_{1}})}})\mathbf{I} \\
\mathbf{T}^{(\mathrm{h_{1}})} & = J^{-1}\mathbf{R}^{e(\mathrm{h_{1}})}\mathbf{M}^{e(\mathrm{h_{1}})}\mathbf{R}^{e(\mathrm{h_{1}})\top}.
\end{aligned}
\end{equation}

\noindent
Meanwhile, the rate of viscoplastic stretching $\mathbf{D}^{p(\mathrm{h_{1}})}$ is coaxial to the driving deviatoric stress, 
\begin{equation}
\mathbf{D}^{p(\mathrm{h_{1}})} =\Dot{\gamma}^{p(\mathrm{h_{1}})}\mathbf{N}^{p(\mathrm{h_{1}})}, \quad \text{with} \quad \mathbf{N}^{p(\mathrm{h_{1}})} = \frac{(\mathbf{M}^{e(\mathrm{h_{1}})})_{0}}{\|(\mathbf{M}^{e(\mathrm{h_{1}})})_{0}\|}.
\end{equation}

\noindent Here, we employ the thermally-activated viscoplasticity model prescribed by,
\begin{equation}
\begin{aligned}
\Dot{\gamma}^{p(\mathrm{h_{1}})} =\Dot{\gamma}_{0}^{\mathrm{h_{1}}}\exp{\left[-\frac{\Delta{G}^{\mathrm{h_{1}}}(1-\tau^{\mathrm{h_{1}}}/s^{\mathrm{h_{1}}})}{k\theta}\right]} \quad \text{where} \quad \tau^{\mathrm{h_{1}}}= \frac{1}{\sqrt{2}}\|(\mathbf{M}^{e(\mathrm{h_{1}})})_{0}\| \\
\end{aligned}
\label{eq:gdh1flowrule}
\end{equation}
\noindent
with the pre-exponential factor $\Dot{\gamma}_{0}^{\mathrm{h_{1}}}$, the activation energy $\Delta{G}^\mathrm{h_{1}}$, Boltzmann's constant $k$, the absolute temperature $\theta$, and the shear strength $s^{\mathrm{h_{1}}}$. Furthermore, to capture the inelastic strain softening beyond yield induced by a reduction of the intermolecular resistance due to rearrangement of the molecular structures during plastic flow in hard domains, a simple saturation-type evolution rule for $s^{\mathrm{h_{1}}}$ is introduced \cite{Boyce1988}, 
\begin{equation}
\Dot{s}^{\mathrm{h_{1}}} =h\left(1-\frac{s^{\mathrm{h_{1}}}}{s_{ss}^{\mathrm{h_{1}}}}\right)\Dot{\gamma}^{p(\mathrm{h_{1}})}
\label{eq:sevolution}
\end{equation}

\noindent
where $h$ is the softening slope. The initial value of shear strength is determined by $s_{0}^{\mathrm{h_{1}}}\approx\frac{0.077}{1-\nu}\mu^{h_{1}}$ ($\nu$ is the initial Poisson's ratio) and the shear resistance decreases until reaching the steady state value $s_{ss}^{\mathrm{h_{1}}}=0.5s_{0}^{\mathrm{h_{1}}}$. 
\noindent Furthermore, the first dissipation inequality in Equation (\ref{eq:dissipation}) for $\alpha=$ h$_{1}$ is satisfied,
\begin{equation}
\mathbf{M}^{e(\mathrm{h_{1}})}_{0}:\mathbf{D}^{p(\mathrm{h_{1}})}=\sqrt{2}\tau^{(\mathrm{h_{1}})}\dot{\gamma}^{p(\mathrm{h_{1}})} \geq 0
\end{equation}
since $\tau^{(\mathrm{h_{1}})}\geq0$ and $\dot{\gamma}^{p(\mathrm{h_{1}})}\geq0$ during the whole loading history. 
 
\paragraph{Network component ($\alpha=\mathrm{h_{2}}$)} \mbox{}\\
\noindent For the equilibrium network elasticity, the rate-independent free energy function is taken from the Arruda Boyce eight chain model \cite{Arruda1993} given by
\begin{equation}
\begin{aligned}
& \psi_{\mathrm{R}}^{(\mathrm{h_{2}})} =\mu^{\mathrm{h_{2}}}(\lambda_{L}^{\mathrm{h_{2}}})^{2}\left[\left(\frac{\Bar{\lambda}^{\mathrm{h_{2}}}}{\lambda_{L}^{\mathrm{h_{2}}}}\right)\beta + \ln{\left(\frac{\beta}{\sinh{\beta}}\right)}\right] \\
\text{where} \quad & \beta = \mathscr{L}^{-1}\left(\frac{\Bar{\lambda}^{\mathrm{h_{2}}}}{\lambda_{L}^{\mathrm{h_{2}}}}\right) \\
\end{aligned}
\label{eq:h2energy}
\end{equation}
\noindent with the shear modulus $\mu^{\mathrm{h_{2}}}$ and the limiting chain extensibility $\lambda_{L}^{\mathrm{h_{2}}}=\sqrt{N^{\mathrm{h_{2}}}}$ where $N^{\mathrm{h_{2}}}$ is the average number of effective chain segments in a polymer chain. Also, $\mathscr{L}^{-1}$ is the inverse Langevin function with $\mathscr{L}(x)=\coth(x)-\dfrac{1}{x}$. Moreover, an average stretch is computed by, 
\begin{equation}
\bar{\lambda}^{\mathrm{h_{2}}} = \sqrt{\frac{\text{tr}\Bar{\textbf{B}}^{(\mathrm{h_{2}})}}{3}}
\label{eq:avgstretch}
\end{equation}
The Cauchy stress is then expressed by,
\begin{equation}
\begin{aligned}
\mathbf{T}^{(\mathrm{h_{2}})} &=\frac{\mu^{\mathrm{h_{2}}}}{3J}\left(\frac{\lambda_{L}^{\mathrm{h_{2}}}}{\Bar{\lambda}^{\mathrm{h_{2}}}}\right)\mathscr{L}^{-1}\left(\frac{\Bar{\lambda}^{\mathrm{h_{2}}}}{\lambda_{L}^{\mathrm{h_{2}}}}\right)\Bar{\mathbf{B}}^{(\mathrm{h_{2}})}_{0} \\
\text{where} \quad \Bar{\mathbf{B}}^{(\mathrm{h_{2}})}_{0} & = \bar{\mathbf{B}}^{(\mathrm{h_{2}})} - \frac{1}{3}(\text{tr}\bar{\mathbf{B}}^{(\mathrm{h_{2}})})\mathbf{I}.
\end{aligned}
\end{equation}

In this exemplar PUU material, the stretch-induced softening (Muillins’ effect) is extensively observed in cyclic deformation experiments. In subsequent reloading cycles, the stress response is remarkably softened. Such a softening phenomenon in this class of two-phase PUU materials was found to be attributed to the microstructural breakdown mainly in the hard domains, which was revealed by \textit{in situ} SAXS and WAXS measurements \cite{Rinaldi2010,Rinaldi2011}.

To capture the Mullins’ effect in the material, a simple damage-based microstructural evolution model is employed, following the simple assumption of chain network alteration (i.e., $\mu^{\mathrm{h_{2}}}N^{\mathrm{h_{2}}}=\text{constant}$) \cite{Marckmann2002, Cho2013a, Mao2017}. The microstructural breakdown in the hard domains results in the increase of the number of effective Kuhn segments $N^{\mathrm{h_{2}}}$ in a polymer chain. Due to the constraint in the network alteration model, the network modulus $\mu^{\mathrm{h_{2}}}$ decreases, culminating in the elastic softening behavior upon deformation. This microscopic damage process due to the network breakdown is irreversible and energy-dissipative. It occurs only when the average stretch $\bar{\lambda}^{\mathrm{h_{2}}}$ is greater than any previously attained upon stretching. To capture this irreversible process, an internal variable $\Bar{\lambda}_{\mathrm{max}}$ is introduced as the maximum average stretch over the loading history,
\begin{equation}
\Bar{\lambda}_{\mathrm{max}} \, \defeq \, \max_{\zeta\in[0,t]}[\Bar{\lambda}^{\mathrm{h_{2}}}(\zeta)].
\label{eq:h2lambdamax}
\end{equation}
The limiting chain extensibility $\lambda_{L}^{\mathrm{h_{2}}}=\sqrt{N^{\mathrm{h_{2}}}}$ evolves irreversibly with the damage variable when the current chain network is stretched above the maximum average stretch imposed over the loading history, i.e., 

\singlespacing
\begin{equation}
\begin{aligned}
& \dot\lambda_{L}^{\mathrm{h_{2}}}=A(\lambda_{L,ss}^{\mathrm{h_{2}}}-\lambda_{L}^{\mathrm{h_{2}}})\,\dot{\Bar\lambda}_{\mathrm{max}} \\ \\
\text{where} \quad & \dot{\Bar\lambda}_{\mathrm{max}}=
\begin{cases}
0, &\Bar{\lambda}^{\mathrm{h_{2}}} < \Bar{\lambda}_{\mathrm{max}} \\ 
\dot{\bar{\lambda}}^{\mathrm{h_{2}}},  & \Bar{\lambda}^{\mathrm{h_{2}}} \geq \Bar{\lambda}_{\mathrm{max}}
\end{cases} \, .
\end{aligned}
\label{eq:h2lockdiff}
\end{equation}
\doublespacing

\bigskip
\noindent
Here $\lambda_{L0}^{\mathrm{h_{2}}}$ is the initial limiting chain extensibility, $\lambda_{L,ss}^{\mathrm{h_{2}}}$ is the saturated limiting chain extensibility, and $A$ is a parameter that controls the rate of damage evolution. The evolving chain extensibility in Equation (\ref{eq:h2lockdiff}) is explicitly given by
\begin{equation}
\lambda_{L}^{\mathrm{h_{2}}} = \lambda_{L,ss}^{\mathrm{h_{2}}}- (\lambda_{L,ss}^{\mathrm{h_{2}}}-\lambda_{L0}^{\mathrm{h_{2}}})\exp{\left(-A\left(\Bar{\lambda}_{\max} - 1 \right)\right)}.
\label{eq:h2lockevolution}
\end{equation}

\noindent Furthermore, to show the dissipation inequality in Equation (\ref{eq:dissipation}) for $\alpha=$ h$_{2}$, we compute,

\begin{equation}
\begin{aligned}
\frac{\partial\psi_{\mathrm{R}}^{(\mathrm{h_{2}})}\left(\mathbf{C}, \Bar{\lambda}_{\max}\right)}{\partial\Bar{\lambda}_{\max}} = & \mu^{\mathrm{h_{2}}}(\lambda_{L}^{\mathrm{h_{2}}})^{2}\frac{\partial}{\partial\Bar{\lambda}_{\max}}\left(\left[\left(\frac{\bar{\lambda}^{\mathrm{h_{2}}}}{\lambda_{L}^{\mathrm{h_{2}}}}\right)\beta + \ln{\left(\frac{\beta}{\sinh{\beta}}\right)}\right]\right) \\
= &\mu^{\mathrm{h_{2}}}(\lambda_{L}^{\mathrm{h_{2}}})^{2}\left(-\frac{\bar{\lambda}^{\mathrm{h_{2}}}}{(\lambda_{L}^{\mathrm{h_{2}}})^{2}}\left(\frac{\partial\lambda_{L}^{\mathrm{h_{2}}}}{\partial\Bar{\lambda}_{\max}}\right)\beta\right) \leq 0.
\end{aligned}
\end{equation}
Since $\dot{\Bar{\lambda}}_{\max} \geq 0$ throughout the loading history, we have,
\begin{equation}
-\frac{\partial\psi_{\mathrm{R}}^{(\mathrm{h_{2}})}\left(\mathbf{C}^{(\mathrm{h_{2}})}, \Bar{\lambda}_{\mathrm{max}}\right)}{\partial\Bar{\lambda}_{\mathrm{max}}}\dot{\Bar{\lambda}}_{\max} \geq 0. \\
\label{eq:NHdissineq}
\end{equation}

\subsubsection{Soft domain}
Since the soft domain is fully relaxed at low strain rate (or long-time scale), the resistance from the soft domain is assumed to be negligible at low strain rate, as revealed by the dynamic mechanical analysis for this PUU material (See Figure 3 in Rinali et al \cite{Rinaldi2010}). However, the soft domain provides additional stiffness at high strain rate (or short-time scale). Also, micro-rheological mechanisms to account for resistances in the soft domain are analogous to those in the hard domain presented in the previous section.
\paragraph{Intermolecular component ($\alpha=\mathrm{s_{1}}$)} \mbox{}\\
\noindent A hyperelastic-viscoplastic mechanism is again employed to capture a time-dependent intermolecular resistance in the soft domains. The constitutive relations for s$_{1}$ are very similar to those in h$_{1}$. However, the inelastic strength $s^{(\mathrm{s_{1}})}$ does not evolve during inelastic flow in the soft domain. In summary,
\begin{itemize}
\item Free energy
\begin{equation}
\psi_{\mathrm{R}}^{(\mathrm{\mathrm{s_{1}}})} =\mu^\mathrm{s_{1}}|{\mathbf{E}^{e(\mathrm{s_{1}})}_{0}}|^{2}
\end{equation}
\item Mandel stress and Cauchy stress
\begin{equation}
\begin{aligned}
\mathbf{M}^{e(\mathrm{s_{1}})} &  = 2\mu^\mathrm{s_{1}}\mathbf{E}^{e(\mathrm{s_{1}})}_{0}\\
\mathbf{T}^{(\mathrm{s_{1}})} & =J^{-1}\mathbf{R}^{e(\mathrm{s_{1}})}(\mathbf{M}^{e(\mathrm{s_{1}})})\mathbf{R}^{e(\mathrm{s_{1}})\mathrm{T}}
\end{aligned}
\end{equation}
\item Flow rule
\begin{equation}
\begin{aligned}
\Dot{\gamma}^{p(\mathrm{s_{1}})} =\Dot{\gamma}_{0}^{\mathrm{s_{1}}}\exp{\left[-\frac{\Delta{G}^{\mathrm{s_{1}}}(1-\tau^{(\mathrm{s_{1}})}/s^{(\mathrm{s_{1}})})}{k\theta}\right]}, \quad \text{where} \quad \tau^{(\mathrm{s_{1}})} = \frac{1}{\sqrt{2}}\|(\mathbf{M}^{e(\mathrm{s_{1}})})_{0}\| \\ \\
\end{aligned}
\label{eq:gds1flowrule}
\end{equation}
\item Dissipation inequality
\begin{equation}
\mathbf{M}^{e(\mathrm{s_{1}})}_{0}:\mathbf{D}^{p(\mathrm{s_{1}})}=\sqrt{2}\tau^{(\mathrm{s_{1}})}\dot{\gamma}^{p(\mathrm{s_{1}})} \geq 0
\end{equation}
\end{itemize}

\paragraph{Network component ($\alpha=\mathrm{s_{2}}$)} \mbox{}\\
Similar to the hard network mechanism (h$_2$), we employ an Arruda-Boyce representation to capture the network resistance in the soft domains.
\begin{itemize}
\item Free energy
\begin{equation}
\begin{aligned}
& \psi_{\mathrm{R}}^{(\mathrm{s_{2}})} =\mu^{\mathrm{s_{2}}}(\lambda_{L}^{\mathrm{s_{2}}})^{2}\left[\left(\frac{\Bar{\lambda}^{e(\mathrm{s_{2}})}}{\lambda_{L}^{\mathrm{s_{2}}}}\right)\beta + \ln{\left(\frac{\beta}{\sinh{\beta}}\right)}\right] \\
\text{where} \quad \quad & \Bar{\lambda}^{e(\mathrm{s_{2}})} = \sqrt{\frac{\text{tr}\Bar{\textbf{C}}^{e(\mathrm{s_{2})}}}{3}} \quad \text{and} \quad \beta = \mathscr{L}^{-1}\left(\frac{\Bar{\lambda}^{e(\mathrm{s_{2}})}}{\lambda_{L}^{\mathrm{s_{2}}}}\right) \\
\end{aligned}
\end{equation}
\item Mandel stress and Cauchy stress
\begin{equation}
\begin{aligned}
\mathbf{M}^{e(\mathrm{s_{2}})} & = \frac{\mu^{\mathrm{s_{2}}}}{3J}\left(\frac{\lambda_{L}^{\mathrm{s_{2}}}}{\Bar{\lambda}^{e(\mathrm{s_{2}})}}\right)\mathscr{L}^{-1}\left(\frac{\Bar{\lambda}^{e(\mathrm{s_{2}})}}{\lambda_{L}^{\mathrm{s_{2}}}}\right)\left[\bar{\mathbf{C}}^{e(\mathrm{s_{2}})}-\frac{1}{3}(\text{tr}\bar{\mathbf{C}}^{e(\mathrm{s_{2}})})\mathbf{I}\right] \\
\mathbf{T}^{(\mathrm{s_{2}})} & =\frac{\mu^{\mathrm{s_{2}}}}{3J}\left(\frac{\lambda_{L}^{\mathrm{s_{2}}}}{\Bar{\lambda}^{e(\mathrm{s_{2}})}}\right)\mathscr{L}^{-1}\left(\frac{\Bar{\lambda}^{e(\mathrm{s_{2}})}}{\lambda_{L}^{\mathrm{s_{2}}}}\right)\Bar{\mathbf{B}}^{e(\mathrm{s_{2}})}_{0}.
\end{aligned}
\end{equation}
\end{itemize}

\bigskip
\noindent Meanwhile, in order to account for the molecular relaxation in the s$_{2}$ mechanism, we employ the viscoelastic flow model proposed by Bergstrom et al., Boyce et al., and Dupaix et al \cite{Bergstrom1998,Llana2000,Dupaix2007}, guided by the reptation theory (Doi and Edwards, 1986). If ambient temperature is sufficiently high (or at long-time scale), 
some of the molecular chains relax from the fully oriented state, resulting in a decrease in the entropic resistance. This reptation-based viscoelastic relaxation in the soft network is given by
\begin{equation}
\mathbf{D}^{p(\mathrm{s_{2}})} =\Dot{\gamma}^{p(\mathrm{s_{2}})}\mathbf{N}^{p(\mathrm{s_{2}})}, \quad \text{where} \quad \mathbf{N}^{p(\mathrm{s_{2}})} = \frac{(\mathbf{M}^{e(\mathrm{s_{2}})})_{0}}{\|(\mathbf{M}^{e(\mathrm{s_{2}})})_{0}\|}.
\end{equation}

\noindent where the viscoelastic strain rate $\dot{\gamma}^{p(\mathrm{s_{2}})}$ is constitutively prescribed by a simple power law,
\begin{equation}
\begin{aligned}
& \Dot{\gamma}^{p(\mathrm{s_{2}})} =C\left(\frac{1}{\bar{\lambda}_{F}-1}\right)\tau^{(\mathrm{s_{2}})m} \\
\text{where} \quad \Bar{\lambda}_{F}= & \sqrt{\frac{\text{tr}\Bar{\mathbf{B}}^{p(\mathrm{s_{2}})}}{3}} \quad \text{and} \quad \tau^{(\mathrm{s_{2}})} = \frac{1}{\sqrt{2}}\|(\mathbf{M}^{e(\mathrm{s_{2}})})_{0}\| 
\end{aligned}
\end{equation}

\bigskip
\noindent
with the relaxation parameter $C$, and the power-law exponent $m$. This viscoelastic flow model can be extended to include the temperature dependence of the molecular relaxation by setting $C=D\exp\left(-Q/R\theta\right)$ with the material parameters $D$, $Q$, and the universal gas constant $R$ \cite{Llana2000,Dupaix2007}. 
\noindent Furthermore, the dissipation inequality in Equation (\ref{eq:dissipation}) for $\alpha=$ s$_{2}$ is satisfied, 
\begin{equation}
\mathbf{M}^{e(\mathrm{s_{2}})}_{0}:\mathbf{D}^{p(\mathrm{s_{2}})}=\sqrt{2}\tau^{(\mathrm{s_{2}})}\dot{\gamma}^{p(\mathrm{s_{2}})} \geq 0
\end{equation}
since $\tau^{(\mathrm{s_{2}})}\geq0$ and $\dot{\gamma}^{p(\mathrm{s_{2}})}\geq0$ during the whole loading history. Each of the material parameters used in the four micro-rheological mechanisms ($\alpha=$ h$_{1}$, h$_{2}$, s$_{1}$, and s$_{2}$) and identification procedures for them are provided in Appendix A.

\subsection{Results: experiment vs. model}
\begin{figure}[b!]
\centering
\hspace*{0.1in}
\includegraphics[width=0.9\textwidth]{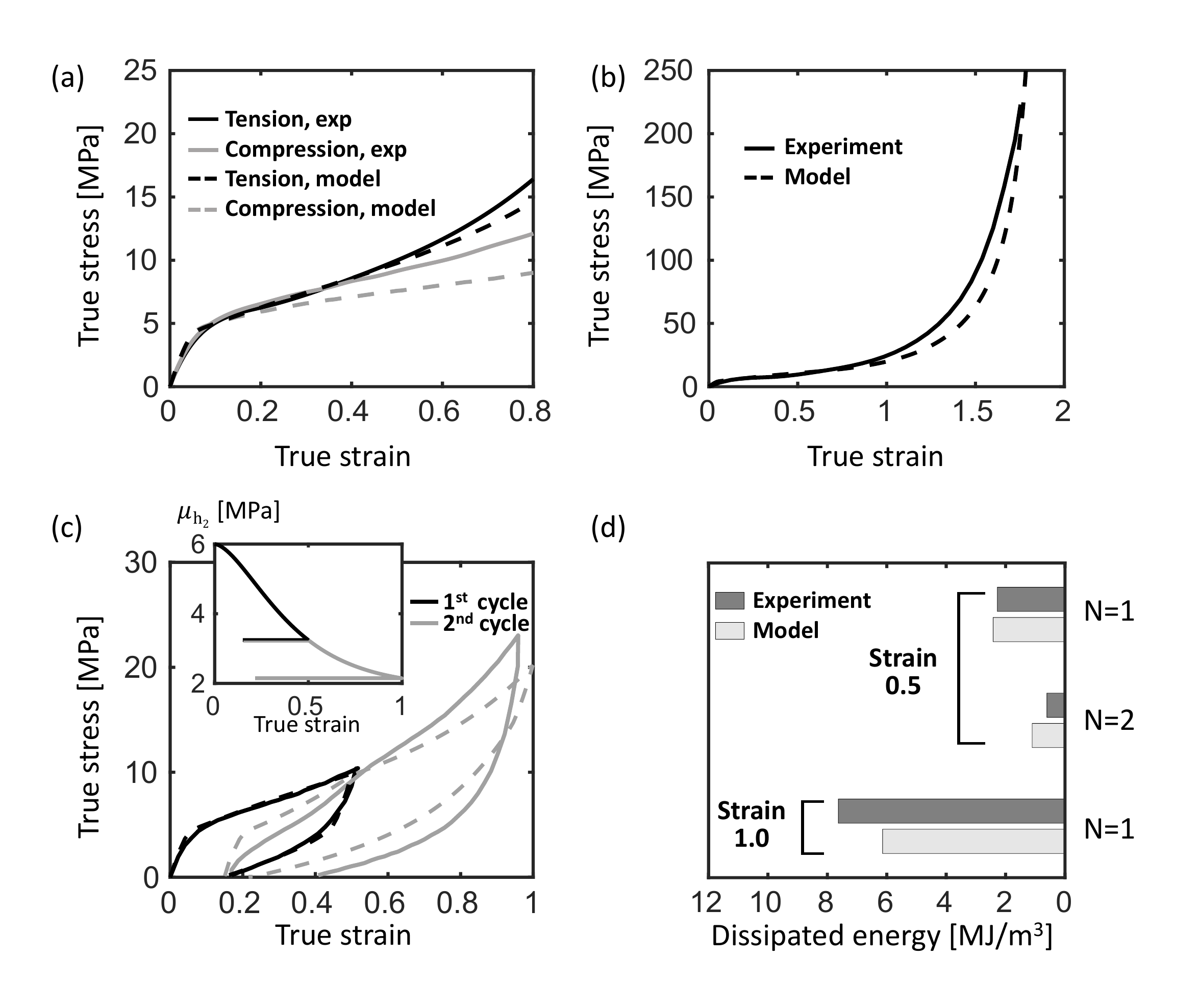} 
\caption{Stress vs. strain curves at a strain rate of 0.01 s$^{-1}$ (solid lines: experiment, dashed lines: model) (a) in tension and compression, (b) in tension up to a true strain of 1.8 and (c) in cyclic tension. (d) Comparison of the dissipated energies for the first cycle (N=1) and the second cycle (N=2) from the hysteresis loop in (c). \\}
\label{fig:result1}
\end{figure}
\noindent Numerically simulated results using the constitutive model are compared against experimental data for the exemplar PUU material under compression and tension over a wide range of strain rates, taken from Rinaldi et al \cite{Rinaldi2010}. Figure \ref{fig:result1} (a) shows the stress-strain behavior in both tension and compression. Our model captures well the overall nonlinear stress-strain behavior including a relatively stiff initial elastic response followed by a yield-like stress rollover and a post-yield hardening due to orientation in the molecular network. Furthermore, the remarkable asymmetry between tension and compression is well captured in the model. This asymmetry is mainly attributed to the greater entropic resistance from the h$_{2}$ mechanism with the greater orientation of the molecular network in tension than in compression. Figure \ref{fig:result1} (b) shows the stress-strain behavior up to a very large tensile true strain of 1.8 ($\sim$ 600\% in stretch). The dramatic stress hardening is excellently captured in the model as the deformation approaches the limiting chain extensibility.

The deformation features under multiple cyclic tensile loading conditions are displayed in Figures \ref{fig:result1} (c) and (d). As shown in Figure \ref{fig:result1} (c), a remarkably softened stress response (Mullins’ effect) in the second cycle is observed in experiments (solid lines). This softening phenomenon is reasonably captured in the model (dashed lines); a decrease in the network modulus (inset of Figure \ref{fig:result1} (c)) in the h$_{2}$ mechanism represents the microstructural breakdown in the hard domains. A quantitative comparison of dissipated energy during cyclic loading and unloading in both experiments and models is displayed in Figure \ref{fig:result1} (d). To do this, we reduced the stress-strain data from both experiments and models to the dissipated work densities in the first and second cycles at increasing strains of 0.5 and 1.0 by integrating the stress-strain curves, following the method described in Greviskes et al \cite{Greviskes2010}. The dissipated work density significantly decreases in the second cycle up to an imposed strain of 0.5 since the stretch-induced softening lessens significantly, which is well captured in the model. However, the dissipated work density at a larger imposed strain of 1.0 in the model is found to be $\sim$ 20\% less than in the experiment; the stress response especially during unloading (dashed gray line) is overpredicted in the model; i.e. while our model captures the overall features (softened stress response and reduced hysteresis in the second cycle) in the cyclic stress-strain behavior of the material reasonably in terms of trend, it overpredicts the stress response during unloading specifically in the second cycle. This indicates that there may be another dissipation mechanism elusive in the material, especially at very large strains. Moreover, the discrepancy during unloading is attributed to unexpected stress relaxation at the end of loading which has been widely observed in the cyclic loading tests for this class of elastomeric materials \cite{Qi2005,Boyce2001}.

\begin{figure}[t!]
\centering
\includegraphics[width=0.9\textwidth]{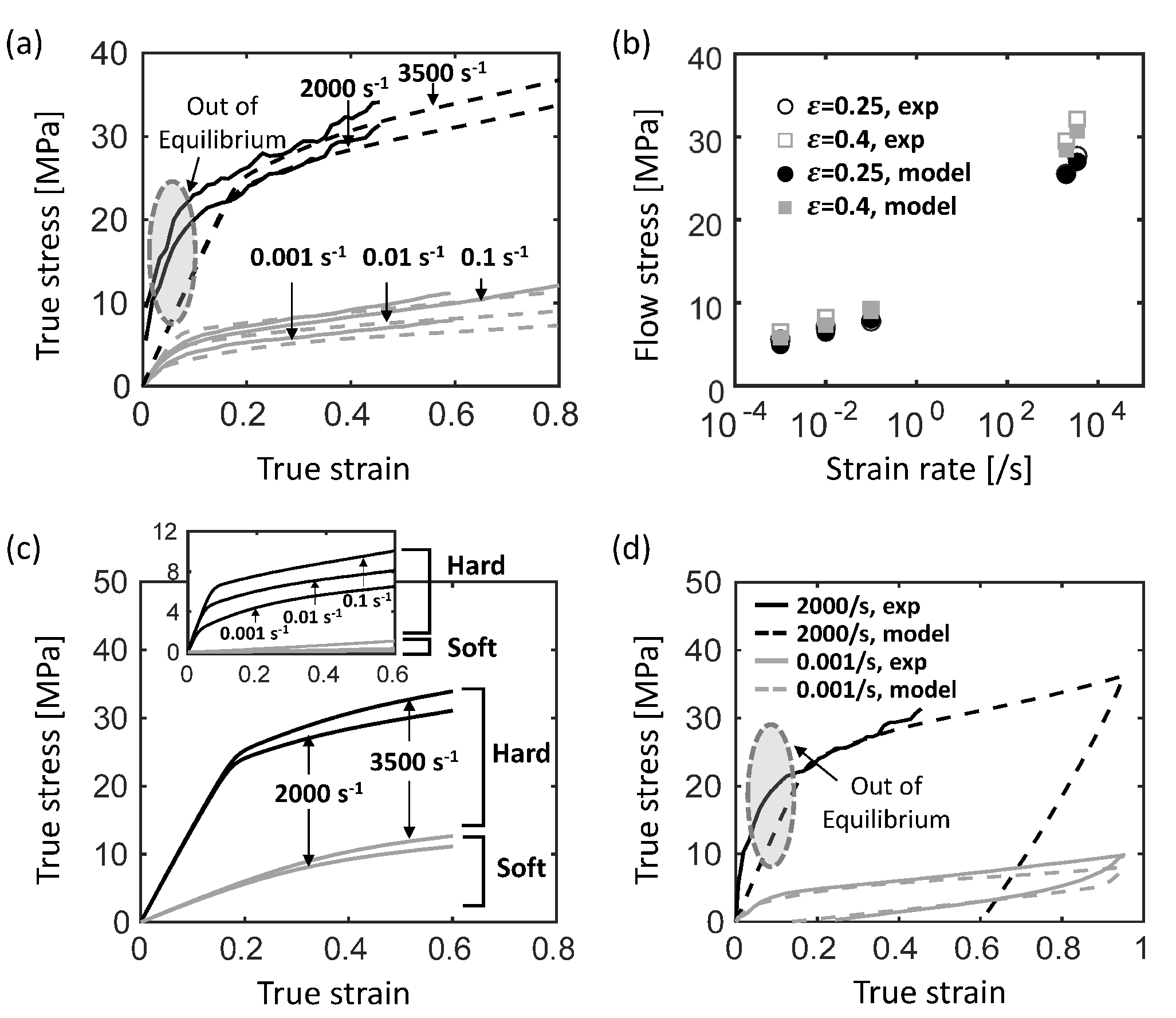} 
\caption{(a) Stress vs. strain curves at low strain rates (0.001, 0.01, and 0.1 s$^{-1}$) to high strain rates (2000 and 3500 s$^{-1}$) in compression (solid lines: experiments, dashed lines: model). (b) Flow stress vs. strain rate at increasing strains of 0.25 and 0.4. (c) Stress contributions from hard (black) and soft (gray) components at low (inset) to high strain rate. (d) Stress vs. strain curves during loading and unloading at strain rates of 0.001 s$^{-1}$ and 2000 s$^{-1}$.}
\label{fig:result2}
\end{figure}

Numerical simulation results in compression at low to high strain rates are displayed along with the experimental data in Figure \ref{fig:result2}. A comparison of the stress-strain curves between models and experiments is shown in Figure \ref{fig:result2} (a) with good agreement at strain rates of 0.001, 0.01, 0.1 and 2000 and 3500 s$^{-1}$. Furthermore, Figure \ref{fig:result2} (b) shows the rate-sensitivity in flow stresses at increasing strains of 0.25 and 0.4 as a function of strain rate for both experiments and models. According to the reduced experimental data (open symbols), there is an apparent change in the rate-sensitivity in the vicinity of a moderate strain rate of $\sim$ 1.0 s$^{-1}$, which is well captured in the model. This transition in the rate-sensitivity is attributed to negligible stress contribution from the soft domains in the material at low strain rates (or at long time scales). To support this, the individual stress contributions from the hard and soft domains are displayed at low to high strain rates in Figure \ref{fig:result2} (c). As shown in the inset of Figure \ref{fig:result2} (c), the stress contribution from the soft domains is found to be negligible at low strain rates of 0.001, 0.01, and 0.1 s$^{-1}$ while it provides a significant, additional stiffness at high strain rates of 2000 and 3500 s$^{-1}$. Lastly, the simulated stress-strain curves during loading and unloading are presented together with those in experiments at strain rates of 0.001 and 2000 s$^{-1}$ (only for loading at 2000 s$^{-1}$) in Figure \ref{fig:result2} (d). At a high strain rate, the mechanical behavior becomes more plastomeric with greater hysteresis and residual strain while it becomes more elastomeric with a decreased dissipation and increased shape recovery at a low strain rate. Our model captures the overall hysteretic behavior traveling between the rubbery (or elastomeric) and the glassy (or plastomeric) polymeric features from low to high strain rates. In summary, our model shows an excellent ability to predict the resilient yet dissipative mechanical behavior of this PUU material from quasi-static ($\sim$ 0.001 s$^{-1}$) to high strain rate ($\sim$ 1000 s$^{-1}$).

\clearpage
\section{Extreme, dynamic behavior of PUU in micro-particle impact}
\subsection{Laser-induced micro-particle impact test}
\begin{figure}[b!]
\centering
\hspace*{-0.2in}
\includegraphics[width=1.1\textwidth]{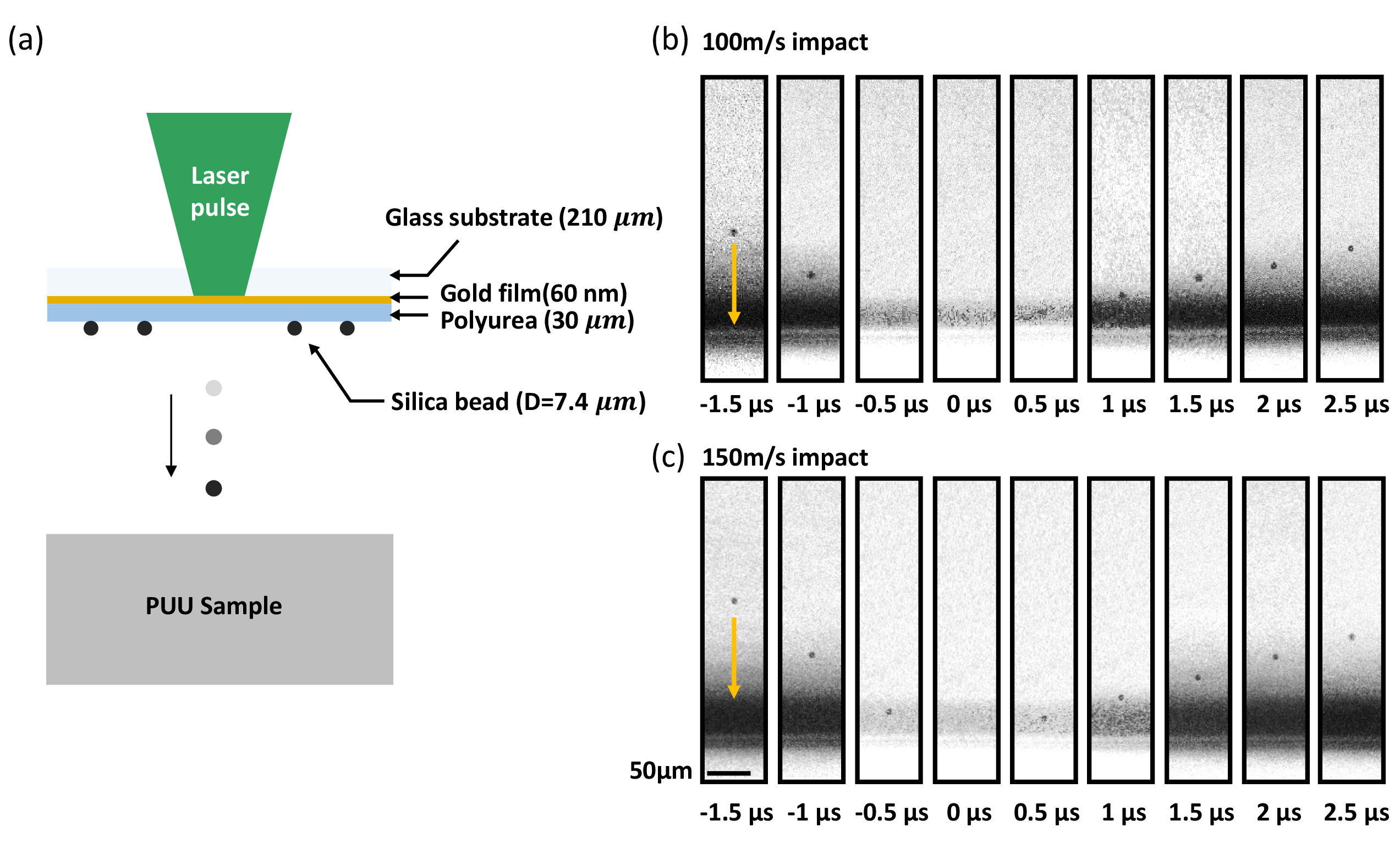} 
\caption{(a) Schematic illustration of the laser-induced particle impact test (LIPIT), and sequential images of the micro-bead trajectories prior to and after impact at (b) 100 m/s and (c) 150 m/s.}
\label{fig:lipit}
\end{figure}
We further address the extreme behavior of the PUU material using the laser-induced particle impact test (LIPIT) platform, by which strongly inhomogeneous deformation in the material at extreme strains and strain rates are elucidated. The LIPIT enables the PUU samples to reach ultrafast strain rates beyond 10$^{6}$ s$^{-1}$ not available in the SHPB setup discussed in Section 2. The LIPIT experimental setup has been recently described \cite{Veysset2016,Mostafa2018,Veysset2017} and used to study the micro-impact behavior of a variety of materials, including nano films, hydrogels, metals, and polymers \cite{Hsieh2021}. In short, as depicted in Figure \ref{fig:lipit} (a), a nanosecond laser pulse (Nd:YAG, 10 ns duration, 532 nm wavelength) is focused onto a launch-pad assembly that consists of a glass substrate, a thin metallic layer (60 nm gold film), and a polyurea layer (30 $\mu$m) on top of which solid silica particles (7.4 $\mu$m diameter) are dispersed. Laser ablation and plasma expansion lead to fast deformation in the polyurea layer, which propels a single silica particle in the air toward the specimen target (here, the exemplar PUU sample), positioned about 750 $\mu$m away, whose surface is normal to the impact direction ($\pm$ 2 degrees). Impact events are recorded using a high-speed camera (SIMX16, Specialized Imaging) with 5-nanosecond exposure time and a diode laser (30$\mu$s duration, 640 nm wavelength, SI-LUX640, Specialized Imaging) for illumination. Figures \ref{fig:lipit} (b) and (c) show two recorded impacts at 100 m/s and 150 m/s with an interframe time of 500 ns. Images are cropped from their original size (400$\times$300 $\mu$m field of view) to focus on the impact site and the image contrast is adjusted in some frames to highlight the impact surface. 

Furthermore, we conducted numerical simulations for the laser-induced micro-particle impact tests of the PUU material using a finite element solver, by which the predictive capabilities of our constitutive model and its numerical implementation are validated for the highly inhomogeneous, extreme deformation event. The constitutive model presented in Section 2 was numerically implemented for use in the finite element solver. All the details regarding an implicit time integration procedure for updating the stress tensor, state variables and kinematic quantities are provided in Appendix B. Moreover, an axisymmetric quadrilateral reduced-integration element was used with a hourglass control scheme. The micro-particle was modeled as a rigid body ($\rho_{bead}=1850$ $kg/m^{3}$); and a frictionless, hard contact constraint was employed between the PUU sample and the bead.

\subsection{Results: experiment vs. simulation}
\begin{figure}[t!]
\centering
\hspace*{-0.4in}
\includegraphics[width=1.1\textwidth]{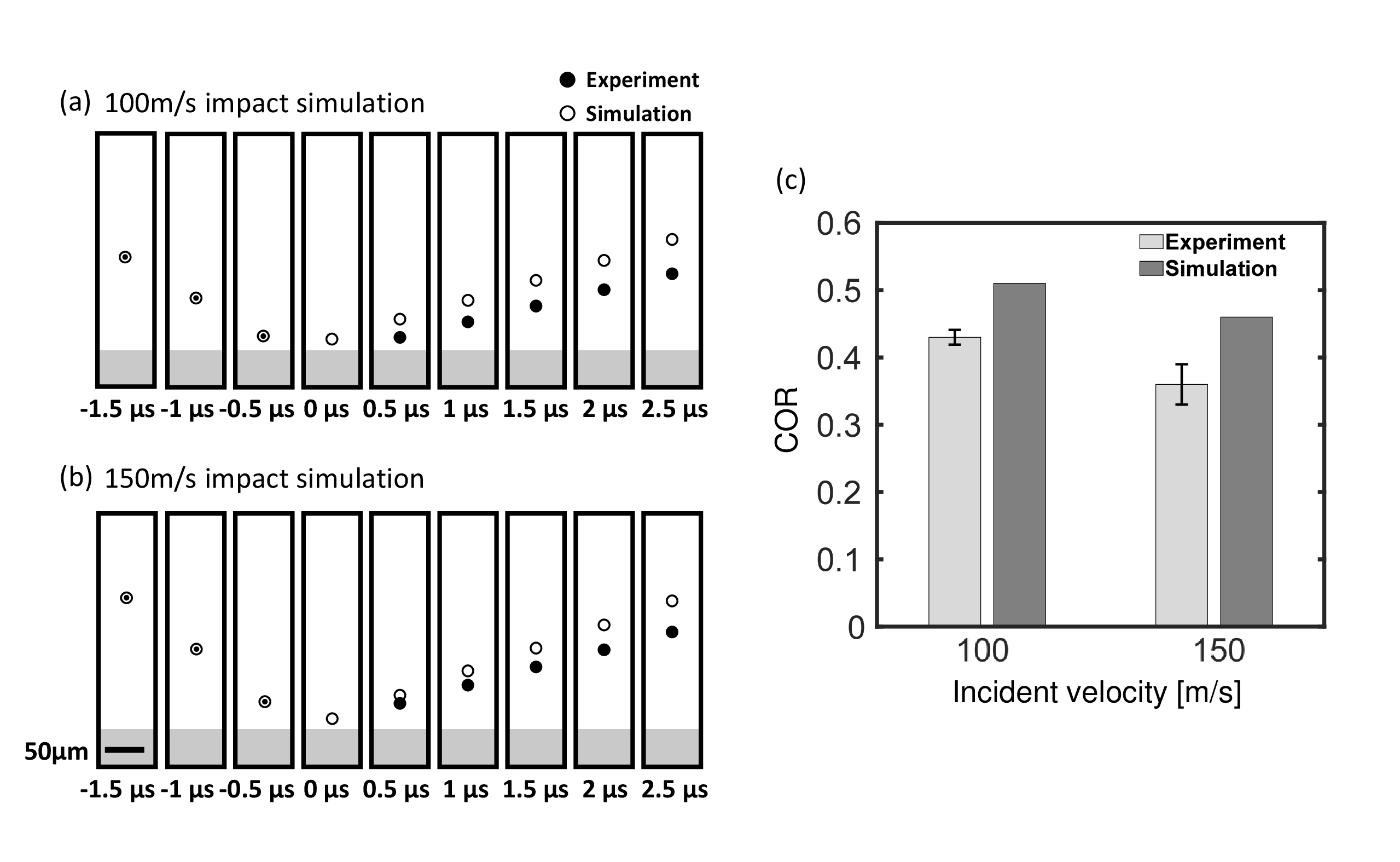} 
\caption{Micro-bead trajectories in experiments and numerical simulations at (a) 100 m/s and (b) 150 m/s. (c) Coefficients of restitution. }
\label{fig:cor}
\end{figure}

Here, we show the results of the LIPIT experiments and compare the major features of the experimental data with those from the numerical simulations. Figures \ref{fig:cor} (a) and (b) show the trajectories of the micro-bead prior to and after impact from both experiments (closed circles) and numerical simulations (open circles) at two different impact velocities. The highly asymmetric behavior between the incident and rebound trajectories is observed in both experiments and numerical simulations, attributed to significant dissipation processes in the PUU sample under impact. The numerical simulations capture the asymmetric trajectories at both velocities reasonably well. The results from the experiments and numerical simulations are further reduced to the coefficients of restitution, COR (defined as the ratio of rebound speed to incident speed) in Figure \ref{fig:cor} (c). As the impact velocity increases, the COR is found to decrease (0.43 $\rightarrow$ 0.36), which is reasonably captured in the numerical simulations (0.51 $\rightarrow$ 0.46). As discussed in Section 2.4, at higher strain rates, hysteresis and energy dissipation were found to increase (e.g. Figure \ref{fig:result2}), which culminates in the decrease in COR with increasing impact velocities. However, in both 100 m/s and 150 m/s, the COR in numerical simulations is found to be $\sim$ 20\% higher than the experimentally measured COR. Although the overall response during impact is well captured in terms of trend, the unloading response in the material after impact is poorly predicted in the numerical simulations, similar to the hysteresis difference between the experimental data and the numerically simulated stress-strain curves at large strains discussed in Section 2.4.

\begin{figure}[b!]
\centering
\hspace*{-0.6in}
\includegraphics[width=1.15\textwidth]{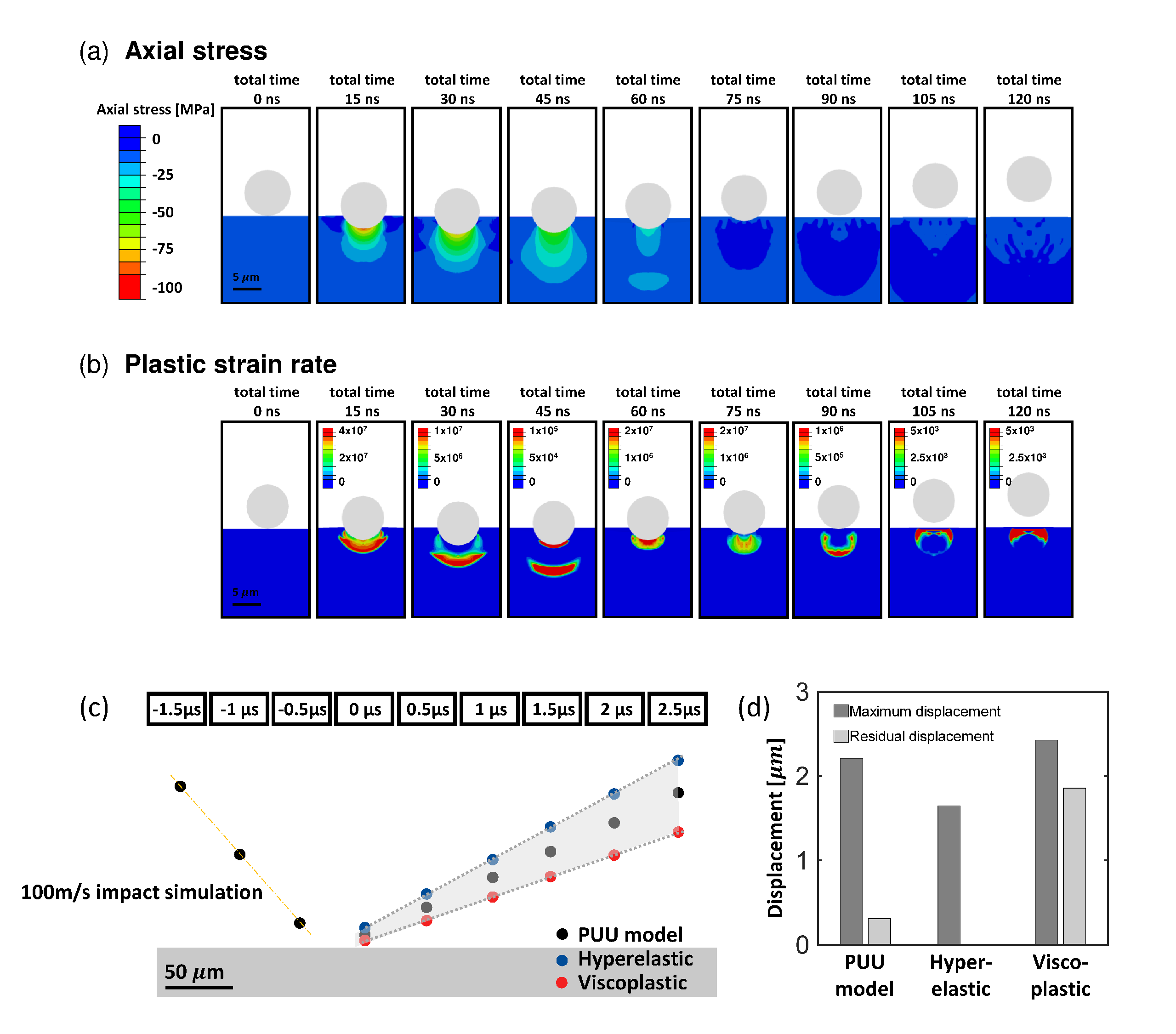} 
\caption{Sequential images of (a) axial stress contours and (b) plastic strain rate contours in numerical simulation (100 m/s). (c) Micro-bead trajectories and (d) maximum and residual deformations in numerical simulations on PUU, hyperelastic and viscoplastic models. }
\label{fig:contour}
\end{figure}

The responses in the PUU sample in the vicinity of the impact surface are further detailed in Figure \ref{fig:contour} on the numerically simulated contours of the deformation field.\footnote{As shown in Figures \ref{fig:lipit} (b) and (c), the trajectories of the micro-bead cannot be identified near the impact surface due to shadow in experiments.} A significantly inhomogeneous deformation field develops with a maximum true axial strain over 1.0 ($> 300\%$) right beneath the impact surface. Figures \ref{fig:contour} (a) and (b) show sequential images of simulated contour plots for the axial stress and the plastic strain rate with a time interval of 15 ns. The sequential plots show that the elastic stress wavefront is followed by extreme plastic deformation ($>10^{6}$ s$^{-1}$). The elastic stress significantly lessens from 100 MPa to a few MPa within tens of nanoseconds after impact while the relatively high plastic strain rate sustained even after the micro-bead rebounded from the impact surface. The significant plastic strain rate after rebound facilitates further shape recovery of the PUU surface beyond the elastic recovery. 
The ability of the constitutive model to capture the energy-dissipative yet resilient behavior during this extreme deformation event is further presented in Figures \ref{fig:contour} (c) and (d). We simulated the impact responses with a purely hyperelastic model and a viscoplastic model simply modified from our PUU constitutive model.\footnote{The purely hyperelastic model was constructed from the PUU model, for which all of the inelastic mechanisms including viscoelasticity, viscoplasticity, and the stretch-induced softening were turned off; the viscoplastic model was constructed again from the PUU model, for which the network resistances for both hard and soft components were excluded.} Figure \ref{fig:contour} (c) shows the simulated trajectories of a micro-bead on (1) hyperelastic, (2) viscoplastic, and (3) PUU samples. The COR is the highest in the purely hyperelastic sample due to the absence of the energy dissipation mechanisms. The COR in the PUU sample is found to be located between those in the purely hyperelastic and the viscoplastic samples. Furthermore, Figure \ref{fig:contour} (d) shows a comparison of the maximum displacement during impact and the residual displacement after rebound for the simulated PUU, hyperelastic and viscoplastic samples. As expected, there is no residual deformation in the hyperelastic sample without any dissipation mechanisms; the PUU sample displays a significant shape recovery ($\sim$ 85\%)\footnote{It should also be noted that we could not find permanent surface indents via scanning electron microscopy following impact due to the shallow nature of the indents in experiments on the PUU sample.} while the viscoplastic sample shows a much poorer shape recovery ($\sim$ 20\%). These results on the energy dissipation (COR) and the shape recovery (residual displacement) in the PUU, hyperelastic and viscoplastic samples strongly supports that, in the PUU material, not only can a significant energy dissipation be achieved from multiple dissipation sources of viscoelasticity, viscoplasticity and the softening but multiple energy storage mechanisms within the hard and soft components enable remarkable elastic-inelastic shape recovery. Our constitutive model nicely captures both resilience and dissipation in the PUU material especially upon extreme deformation by taking advantages of both hyperelastic and viscoplastic constitutive laws.

\clearpage
\section{Conclusion}
In this work, the large strain behavior of a polyurethane-urea material from low to extreme strain rates has been addressed. We proposed a microstructurally- and thermodynamically based constitutive model capable of capturing key elastic and inelastic features of the representative PUU material. We employed multiple micro-rheological representations to account for distinct resistances from hard and soft components in the material. Our model was able to capture the hybrid elastomeric- and plastomeric features attributed to the presence of the distinct hard and soft domains in the material. The predictive capabilities of the model and its numerical implementation have been validated by comparing the experimental data with the results from simulations under large strain tension and compression over a wide range of strains and strain rates. Furthermore, we have elucidated the hybrid features of the PUU material at extreme strains and strain rates using the laser-induced micro-particle impact test. The model captured the main features in resilience and dissipation for a strongly inhomogeneous deformation at an extreme strain rate ($>10^{6}$ s$^{-1}$) reasonably well. Our model should be useful for simulating extreme deformation behaviors of other multi-phase elastomeric or plastomeric materials. 

As previously noted, although the resilient yet dissipative deformation features were well captured in the model, the unloading behaviors involving hysteresis and rebound were not predicted precisely. In the future, additional dissipation mechanisms in this material should be further addressed for a better understanding of the highly nonlinear unloading behavior, especially at extreme strain rates. Moreover, a coupled thermo-mechanical behavior at extreme strains and strain rates is an open area for this important class of two-phase soft materials.

\section*{Acknowledgements}
We gratefully acknowledge valuable discussions with Mary Boyce and the financial support provided by the National Research Foundation of Korea (Grant No. 2020R1C1C101324813).

\clearpage
\renewcommand*\appendixpagename{Appendix}
\renewcommand*\appendixtocname{Appendix}
\begin{appendices}
\section{Material parameters}
Material parameters used in the model are given in Table \ref{Tab:parameter}.
\begin{table}[h!]
\doublespacing
\centering
\begin{tabular}{ll|c|c|c|c}
\hline
\textbf{Elastic-viscoplastic}   &  &  &  & $\alpha=$ h$_{1}$ & $\alpha=$ s$_{1}$ \\ \hline
\multicolumn{2}{l|}{Stress}  & $\mu^{\alpha}_{0}$ & {[}MPa{]} & 25 & 13            \\
\multicolumn{2}{l|}{\fontsize{10pt}{10pt}\selectfont{$\mathbf{T}^{(\alpha)} = J^{-1}\mathbf{R}^{e(\alpha)}\left(2\mu^\alpha\mathbf{E}^{e(\alpha)}_{0}+K(\mathrm{tr}\mathbf{E^{e(\alpha)}})\mathbf{I}\right)\mathbf{R}^{e(\alpha)\top}$}} & $K$ & {[}GPa{]} & 1.5 & N/A \\
\multicolumn{2}{l|}{Flow rule} & $\Delta{G}^{\alpha}$ & {[}$10^{-20}$J{]} & 2.7 & 0.25 \\
\multicolumn{2}{l|}{\fontsize{10pt}{10pt}\selectfont{$\Dot{\gamma}^{p(\alpha)} =\Dot{\gamma}_{0}^{\alpha}\exp{\left[-\frac{\Delta{G}^{\alpha}(1-\tau^{(\alpha)}/s^{(\alpha)})}{k\theta}\right]}$}}  & $\Dot{\gamma_{0}}^{\alpha}$ & {[}$s^{-1}${]} & 0.06 & 353            \\
\multicolumn{2}{l|}{Softening} & $s_{0}^{\alpha}$ & {[}MPa{]} & 2.5 & 1.5 \\
\multicolumn{2}{l|}{\fontsize{10pt}{10pt}\selectfont{$\Dot{s}^{\alpha} =h\left(1-\frac{s^{\alpha}}{s_{ss}^{\alpha}}\right)\Dot{\gamma}^{p(\alpha)}$}} & $h$ & {[}MPa{]} & 5 & N/A \\ [2ex] \hline 
\textbf{Hyperelastic}    &  &  &  & $\alpha=$ h$_{2}$ & $\alpha=$ s$_{2}$ \\ \hline
\multicolumn{2}{l|}{Stress} & $\mu^{\alpha}$ & {[}MPa{]} & 6 & 4 \\
\multicolumn{2}{l|}{\fontsize{10pt}{10pt}\selectfont{$\mathbf{T}^{(\alpha)} =\frac{\mu^{\alpha}}{3J}\left(\frac{\lambda_{L}^{\alpha}}{\Bar{\lambda}^{\alpha}}\right)\mathscr{L}^{-1}\left(\frac{\Bar{\lambda}^{\alpha}}{\lambda_{L}^{\alpha}}\right)\Bar{\mathbf{B}}^{(\alpha)}_{0}$}} & $\lambda_{L0}^{\alpha}$ &  & $\sqrt{4.5}$ & $\sqrt{20}$  \\
\multicolumn{2}{l|}{Softening}  & $\lambda_{L,ss}^{\alpha}/\lambda_{L0}^{\alpha}$ & &  $\sqrt{3}$ & N/A \\
\multicolumn{2}{l|}{\fontsize{10pt}{10pt}\selectfont{$\lambda_{L}^{\alpha} = \lambda_{L,ss}^{\alpha}- (\lambda_{L,ss}^{\alpha}-\lambda_{L0}^{\alpha})\exp{\left(-A\left(\Bar{\lambda}_{\max} - 1 \right)\right)}$}}  & $A$ & & 3 & N/A \\
\multicolumn{2}{l|}{Molecular relaxation} & $C$ & {[}Pa$^{-1}$s$^{-1}${]} & N/A & $9\times10^{-12}$ \\
\multicolumn{2}{l|}{\fontsize{10pt}{10pt}\selectfont{$\Dot{\gamma}^{p(\alpha)} =C\left(\frac{1}{\bar{\lambda}_{F}-1}\right)\tau^{(\alpha)m}$}}    & $m$ & & N/A & 1.5 \\ [2ex] \hline
\multicolumn{6}{r}{\small{N/A: Not applicable}}
\end{tabular}
\caption{Material parameters used in the constitutive model ($\rho_{\mathrm{PUU}}=1100$ $kg/m^{3}$).}
\label{Tab:parameter}
\end{table}

\noindent Here, a procedure to identify the material parameters is briefly presented.
\subsection{Hard domain}
\begin{itemize}
\item We assumed that the stress contribution from the soft domains is negligible in quasi-static strain rates since the soft domains are completely relaxed. Thus, the material parameters for the hard mechanisms are determined using the low strain rate data. The initial Young's modulus was found to be $E_{\mathrm{hard}}=90$ MPa from the stress-strain curve at a low strain rate of 0.01 s$^{-1}$. The elastic contributions from the h$_{1}$ and h$_{2}$ mechanisms are identified using the difference in tension and compression data since the asymmetry in tension and compression mainly results from the entropic resistance in the h$_{2}$ mechanism. The initial shear modulus of the h$_{1}$ and h$_{2}$ mechanisms were found to be $\mu^{\mathrm{h_{1}}}=25$ MPa and $\mu^{\mathrm{h_{2}}}=6$ MPa, respectively.

\item The material parameters associated with viscoplastic flow in the h$_{1}$ mechanism was identified using the shear yield stress $\tau^{(\mathrm{h_{1}})}$ vs. shear strain rate $\dot\gamma^{p(\mathrm{h_{1}})}$ plot. From the thermally-activated viscoplastic flow model in Equation (\ref{eq:gdh1flowrule}), the equation can be rewritten as $\tau^{(\mathrm{h_{1}})} =A^{\mathrm{h_{1}}}\ln\dot\gamma^{p(\mathrm{h_{1}})}+B^{\mathrm{h_{1}}}$ where $A^{\mathrm{h_{1}}} = s^{\mathrm{h_{1}}}\left(\frac{k\Theta}{\Delta G^{\mathrm{h_{1}}}}\right)$ and $B^{\mathrm{h_{1}}}=s^{\mathrm{h_{1}}}-A^{\mathrm{h_{1}}}\ln\dot\gamma_{0}^{\mathrm{h_{1}}}$. The initial shear strength was assumed to be $s_{0}^{\mathrm{h_{1}}}=2.5$ MPa using the relation $s=\frac{0.077\mu}{1-\nu}$ predicted for glassy polymers (the Poisson's ratio $\nu=0.49$). Hence, the least-square fit of the equation provides $\Delta G^{\mathrm{h_{1}}} = 2.7\times10^{-20}$J and $\dot\gamma_{0}^{\mathrm{h_{1}}}=0.06 \,\mathrm{s}^{-1}$.

\item We then identified the saturated value of chain extensibility $\lambda_{L,ss}^{\mathrm{h_{2}}}$ from the stress-strain curve in tension with a stress-upturn. As shown in Figure \ref{fig:result1} (b), the limiting chain extensibility was found to be $\lambda_{L,ss}^{\mathrm{h_{2}}}=\sqrt{\frac{\lambda_{1}^{2}+\lambda_{2}^{2}+\lambda_{3}^{2}}{3}}=\sqrt{\frac{1}{3}(e^{2\epsilon}+2e^{-2\epsilon})}=\sqrt{14}$ where $\epsilon$ is the maximum logarithmic strain. We obtained the rate of the initial shear modulus to the saturated shear modulus $\mu^{\mathrm{h_{2}}}_{0}/\mu^{\mathrm{h_{2}}}_{ss}=3$ from the stretch-induced softening in cyclic tension data in Figure \ref{fig:result1} (c). The ratio of the saturated chain extensibility to the initial chain extensibility was identified with $\lambda_{L,ss}^{\mathrm{h_{2}}}/\lambda_{L,0}^{\mathrm{h_{2}}}=\sqrt{N_{L,ss}^{\mathrm{h_{2}}}/N_{L,0}^{\mathrm{h_{2}}}}=\sqrt{3}$ using the assumption of chain network alteration $\mu^{\mathrm{h_{2}}}N^{\mathrm{h_{2}}}=\text{constant}$. Therefore, we obtained the initial chain extensibility $\lambda_{L,0}^{\mathrm{h_{2}}}=\sqrt{4.5}$. The parameter $A$ that controls the rate of the softening was found to be $\sim$3.
\end{itemize}

\subsection{Soft domain}
\begin{itemize}
\item The initial Young's modulus of the soft domain was found to be $E_{\mathrm{soft}}=E_{\mathrm{total}}-E_{\mathrm{hard}}=55$ MPa from the stress-strain curve at high strain rate. We assumed that the ratio between the elastic contributions from the $\mathrm{s_{1}}$ and $\mathrm{s_{2}}$ mechanisms is similar to that of the hard domains. Therefore, the shear moduli were identified to be $\mu^{\mathrm{s_{1}}}=13$ MPa and $\mu^{\mathrm{s_{2}}}=4$ MPa, respectively.

\item The material parameters for the viscoplastic response in the s$_{1}$ mechanism were identified from the least-square fit of $\tau^{(\mathrm{s_{1}})} =A^{\mathrm{s_{1}}}\ln\dot\gamma^{p(\mathrm{s_{1}})}+B^{\mathrm{s_{1}}}$ where $A^{\mathrm{s_{1}}} = A^{\mathrm{total}}-A^{\mathrm{h_{1}}}$ and $B^{\mathrm{s_{1}}}=B^{\mathrm{total}}-B^{\mathrm{h_{1}}}$. Here, $A^{\mathrm{total}}$ and $B^{\mathrm{total}}$ were identified from the shear yield stress vs. shear strain rate plot of the high strain rate data. We then obtained $\Delta G^{\mathrm{s_{1}}} = 2.5\times10^{-21}$J and $\dot\gamma_{0}^{\mathrm{s_{1}}}=353 \,\mathrm{s}^{-1}$.

\item The material parameters in the s$_{2}$ mechanism were identified by subtracting the extrapolated stress responses of the h$_{1}$, h$_{2}$, and s$_{1}$ mechanisms from the total stress data. The chain extensibility in the s$_{2}$ mechanism was assumed to be $\lambda_{L}^{\mathrm{s_{2}}}\sim\sqrt{20}$ and the material parameters associated with the molecular relaxation were identified to be $C=9\times10^{-12} \text{Pa}^{-1}\text{s}^{-1}$ and $m=1.5$, following the procedures for the molecular relaxation-based viscoelastic mechanisms in Bergstr\"om et al., Boyce et al., and Dupaix et al \cite{Bergstrom1998,Llana2000,Dupaix2007}.
\end{itemize}
\bigskip

\section{Time integration procedure}
We present a time integration procedure for the time-dependent viscoelastic or viscoplastic mechanisms (h$_{1}$, s$_{1}$, s$_{2}$) in our constitutive model, following the algorithms in Weber and Anand \cite{Weber1990a} and Chester \cite{Chesterphd}. The elastic response of the s$_{2}$ component is assumed to be linear elastic, i.e., 
$\mathbf{M}^{e(\mathrm{s_{2}})}=2\mu^{\mathrm{s_{2}}}\mathbf{E}^{e(\mathrm{s_{2}})}_{0}$ to simplify the numerical procedure and increase robustness.

\begin{itemize}
\item Given : $\{\mathbf{F}_{n},\mathbf{F}_{n+1},\mathbf{F}^{p}_{n}, \mathbf{T}_{n}, s_{n}, \dot{\gamma}^{p}_{n}\}$ at time $t_{n}$
\item Calculate : $\{\mathbf{T}_{n+1}, s_{n+1}, \dot{\gamma}^{p}_{n+1}\}$ at time $t_{n+1}$
\end{itemize}

\noindent Using an exponential map for integrating $\dot{\mathbf{F}}^{p}=\mathbf{D}^{p}\mathbf{F}^{p}$, the plastic deformation gradient is calculated,
\begin{equation}
\mathbf{F}^{p}_{n+1} =\exp\left(\Delta t\mathbf{D}^{p}_{n+1}\right)\mathbf{F}^{p}_{n}.
\end{equation}
\noindent Then, the elastic deformation gradient is obtained, 
\begin{equation}
\begin{aligned}
\mathbf{F}^{e}_{n+1}  & = \mathbf{F}_{n+1}\mathbf{F}^{p-1}_{n+1} \\
& =\mathbf{F}_{n+1}\mathbf{F}^{p-1}_{n}\exp\left(-\Delta t\mathbf{D}^{p}_{n+1}\right) \\
& =\mathbf{F}^{e}_{\text{trial}}\exp\left(-\Delta t\mathbf{D}^{p}_{n+1}\right) \qquad \qquad \text{where} \qquad \mathbf{F}^{e}_{\text{trial}} & \defeq\mathbf{F}_{n+1}\mathbf{F}^{p-1}_{n}.
\end{aligned}
\end{equation}

\noindent The polar decomposition of the trial elastic deformation gradient is given with an approximation to simplify the procedure,
\begin{equation}
\begin{aligned}
\mathbf{F}^{e}_{\text{trial}} & =\mathbf{R}^{e}_{\text{trial}}\mathbf{U}^{e}_{\text{trial}} \\
\mathbf{R}^{e}_{\text{trial}} & \approx \mathbf{R}^{e}_{n+1} \\
\mathbf{U}^{e}_{\text{trial}} & \approx \mathbf{U}^{e}_{n+1}\exp\left(\Delta t\mathbf{D}^{p}_{n+1}\right).
\end{aligned}
\end{equation}
\noindent We then obtain the logarithmic elastic strain,
\begin{equation}
\begin{aligned}
\mathbf{E}^{e}_{n+1} & = \ln\left(\mathbf{U}^{e}_{\text{trial}}\exp\left(-\Delta t\mathbf{D}^{p}_{n+1}\right)\right) \\
& = \mathbf{E}^{e}_{\text{trial}}-\Delta t\mathbf{D}^{p}_{n+1} \qquad \qquad \qquad \text{where} \qquad \mathbf{E}^{e}_{\text{trial}} & =\ln\mathbf{U}^{e}_{\text{trial}}.
\end{aligned}
\end{equation}

\noindent Then, the relationship between the trial Mandel stress and the Mandel stress at time $t_{n+1}$ is obtained using the linear elasticity with the logarithmic strain measure, 
\begin{equation}
\begin{aligned}
\mathbf{M}^{e}_{n+1} & =2\mu\left(\mathbf{E}^{e}_{n+1}\right)_{0}+K\text{tr}(\mathbf{E}^{e}_{n+1})\mathbf{I} \\
& = 2\mu\left(\mathbf{E}^{e}_{\text{trial}}\right)_{0} -2\mu\Delta t\mathbf{D}^{p}_{n+1} + K\text{tr}(\mathbf{E}^{e}_{\text{trial}})\mathbf{I} \\
& = \mathbf{M}^{e}_{\text{trial}}-2\mu\Delta t\mathbf{D}^{p}_{n+1} \\
\text{where} \qquad \mathbf{M}^{e}_{\text{trial}} & =2\mu\left(\mathbf{E}^{e}_{\text{trial}}\right)_{0}+K\text{tr}(\mathbf{E}^{e}_{\text{trial}})\mathbf{I} \\
\end{aligned}
\end{equation}

\noindent using $\text{tr}(\mathbf{E}^{e}_{n+1})=\text{tr}(\mathbf{E}^{e}_{\text{trial}})$ since $\mathbf{D}^{p}_{n+1}$ is deviatoric. Then, we have 
\begin{equation}
\left(\mathbf{M}^{e}_{n+1}\right)_{0}  = \left(\mathbf{M}^{e}_{\text{trial}}\right)_{0}-2\mu\Delta t\mathbf{D}^{p}_{n+1}.
\end{equation}

\noindent By defining
\begin{equation}
\begin{aligned}
& \mathbf{D}^{p}_{n+1} =\dot{\gamma}^{p}_{n+1}\mathbf{N}^{p}_{n+1} & \\
& \bar{\tau}_{n+1} =\frac{1}{\sqrt{2}}\left\|\left(\textbf{M}^{e}_{n+1}\right)_0\right\| \qquad & \text{and} \qquad  \mathbf{N}^{p}_{n+1}=\frac{\left(\mathbf{M}^{e}_{n+1}\right)_{0}}{\sqrt{2}\bar{\tau}_{n+1}}\\
& \bar{\tau}_{\text{trial}} =\frac{1}{\sqrt{2}}\left\|\left(\textbf{M}^{e}_{\text{trial}}\right)_0\right\| \qquad & \text{and} \qquad  \mathbf{N}^{p}_{\text{trial}}=\frac{\left(\mathbf{M}^{e}_{\text{trial}}\right)_{0}}{\sqrt{2}\bar{\tau}_{\text{trial}}}, \\
\end{aligned}
\end{equation}
we obtain 
\begin{equation}
\bar{\tau}_{n+1}\mathbf{N}^{p}_{n+1}-\bar{\tau}_{\text{trial}}\mathbf{N}^{p}_{\text{trial}}+\sqrt{2}\mu\Delta t\dot{\gamma}^{p}_{n+1}\mathbf{N}^{p}_{n+1}=0.
\end{equation}

\noindent Then, we arrive at the following important results,
\begin{equation}
\begin{aligned}
& \mathbf{N}^{p}_{n+1}=\mathbf{N}^{p}_{\text{trial}} \\
& \bar{\tau}_{n+1}-\bar{\tau}_{\text{trial}}+\sqrt{2}\mu\Delta t\dot{\gamma}^{p}_{n+1}=0.
\end{aligned}
\end{equation} 

\noindent We solve the implicit equation for $\dot{\gamma}^{p}_{n+1}$ with the flow model from Equation (\ref{eq:gdh1flowrule}),
\begin{equation}
s_{n+1}\left(\frac{k_{b}\theta}{\Delta G}\ln\frac{\dot{\gamma}^{p}_{n+1}}{\dot{\gamma}_{0}}+1\right)-\bar{\tau}_{\text{trial}}+\mu\Delta t\dot{\gamma}^{p}_{n+1}=0.  \\
\end{equation} 
Here, $s_{n+1}$ is obtained using the Euler backward time integration,
\begin{equation}
s_{n+1} = s_{n} + \Delta t\dot{s}_{n+1} \qquad \text{where} \qquad \dot{s}_{n+1}=h\left(1-\frac{s_{n+1}}{s_{ss}}\right)\dot{\gamma}^{p}_{n+1} \\
\end{equation}
Then
\begin{equation}
s_{n+1} = \frac{s_{n}+\Delta th\dot{\gamma}^{p}_{n+1}}{1+\Delta th\dot{\gamma}^{p}_{n+1}/s_{ss}}.
\end{equation}
The implicit equation for $\dot{\gamma}^{p}_{n+1}$ is rewritten in,
\begin{equation}
\left(\frac{s_{n}+\Delta th\dot{\gamma}^{p}_{n+1}}{1+\Delta th\dot{\gamma}^{p}_{n+1}/s_{ss}}\right)\left(\frac{k_{b}\theta}{\Delta G}\ln\frac{\dot{\gamma}^{p}_{n+1}}{\dot{\gamma}_{0}}+1\right)-\frac{1}{\sqrt{2}}\left\|2\mu\left(\mathbf{E}^{e}_{\text{trial}}\right)_{0}\right\|+\mu\Delta t\dot{\gamma}^{p}_{n+1}=0.
\label{eq:nonlinearimp}
\end{equation} 

\noindent The implicit equation in Equation (\ref{eq:nonlinearimp}) is then solved using a combination of bisection and Newton algorithms \cite{NumericalRecipe}. We finally obtain the plastic deformation gradient, $\mathbf{F}^{p}_{n+1}$ at time $t_{n+1}$. Then, the Cauchy stress can be obtained with the elastic deformation gradient $\mathbf{F}^{e}_{n+1}=\mathbf{F}_{n+1}\mathbf{F}^{p-1}_{n+1}$ at time $t_{n+1}$.

\end{appendices}

\clearpage
\singlespacing
\renewcommand*{\bibfont}{\normalfont\fontsize{10pt}{10pt}\selectfont}
\printbibliography
\end{document}